\documentclass[preprint,jcp,floats,amsmath,amssymb]{revtex4}

\usepackage{graphicx}
\usepackage{amsmath}
\usepackage{amssymb}
\usepackage[english]{babel}
\usepackage{MnSymbol}
\usepackage{color}
\usepackage{chemarr}

\begin{document}
\title{Comparison of different moment-closure approximations for \\ stochastic chemical kinetics}
\author{David Schnoerr $^{1,2}$,  Guido Sanguinetti $^{2}$, Ramon Grima $^{1}$}
\affiliation{$^{1}$ School of Biological Sciences, University of Edinburgh, UK \\ $^{2}$ School of Informatics, University of Edinburgh, UK}

\begin{abstract}
In recent years moment-closure approximations (MA) of the chemical master equation have become a popular method for the study of stochastic effects in chemical reaction systems. Several different MA methods have been proposed and applied in the literature, but it remains unclear how they perform with respect to each other. In this paper we study the normal, Poisson, log-normal  and central-moment-neglect MAs by applying them to understand the stochastic properties of chemical systems whose deterministic rate equations show the properties of bistability, ultrasensitivity and oscillatory behaviour. Our results suggest that the normal MA is favourable over the other studied MAs. In particular we found that (i) the size of the region of parameter space where a closure gives physically meaningful results, e.g. positive mean and variance, is considerably larger for the normal closure than for the other three closures; (ii) the accuracy of the predictions of the four closures (relative to simulations using the stochastic simulation algorithm) is comparable in those regions of parameter space where all closures give physically meaningful results; (iii) the Poisson and log-normal MAs are not uniquely defined for systems involving conservation laws in molecule numbers. We also describe the new software package MOCA which enables the automated numerical analysis of various MA methods in a graphical user interface and which was used to perform the comparative analysis presented in this paper. MOCA  allows the user to develop novel closure methods and can treat polynomial, non-polynomial, as well as time-dependent propensity functions, thus being applicable to virtually any chemical reaction system.
\end{abstract}

\maketitle

\section{Introduction}

Biochemical reactions systems frequently comprise species with low copy numbers of molecules which leads to strong stochastic effects \cite{Grima2008}. Under well-mixed and dilute conditions, the chemical master equation (CME) is the accepted description of the dynamics of such systems \cite{Gillespie2007}. For all but the most simple systems, however, no analytic solutions of the CME are known. The standard approach in this case is to use the stochastic simulation algorithm (SSA \cite{Gillespie1977}), a popular Monte-Carlo method that samples from the solution of the CME. However, the SSA is computationally expensive and becomes infeasible for all but the smallest systems, in particular if some of the species occur in high molecule numbers with many reactions happening per unit time. While the derivation of a reduced CME enforcing time scale separation may help in some cases \cite{Gomez2008, Gillespie2009b}, analytical approximations are still an important alternative for the exploration of chemical systems. 

Using the CME one can derive ordinary differential equations for the moments of the numbers of molecules of each species in the system. In general the equation for a given moment is coupled to the equations of higher order moments giving rise to an infinite hierarchy of equations which cannot be solved \cite{McQuarrie1967}. A popular method to approximate the moments of the CME are moment-closure approximations (MA) \cite{Grima2012,Ferm2008,Ullah2009,Verghese2007,Ale2013}. The latter usually express moments above a certain order in terms of lower order moments, thereby closing the moment equations which can then be solved either analytically or numerically. Several different moment-closure methods have been proposed in the literature. The most popular is the normal MA  (also called ``cumulant neglect MA"), which sets all cumulants above a certain order to zero, thus corresponding to a normal distribution \cite{Grima2012,Ferm2008,Ullah2009,Verghese2007,Ale2013}. If all cumulants above order $M$ are set to zero we speak of the ``normal MA of order $M$''. Several other MAs have been proposed to close the moment equations; some common types are the Poisson MA \cite{Nasell2003}, the log-normal MA \cite{Keeling2000} and the central-moment-neglect MA (CMN-MA) \cite{Hespanha2008}.

The purpose of this paper is twofold: (i) an empirical comparison of the predictions of different types of MAs when applied to chemical reaction systems, and (ii) the presentation of a new user-friendly software package which enables the automatic derivation and analysis of MAs.

MAs are an ad-hoc approximation and there is no straightforward way to predict their accuracy. While several different MA methods have been proposed \cite{Ferm2008,Ullah2009,Verghese2007, Nasell2003,Keeling2000,Singh2006,Lakatos2015} and successfully applied \cite{Milner2013, Zechner2012} in the literature, there are few studies analysing and comparing their performance. In \cite{Hespanha2011} the log-normal MA was found to be more accurate than the normal MA for a gene cascade network for one parameter set. In \cite{Grima2012} the accuracy of the normal MA has been investigated for general monostable systems in the limit of large volumes using the system size expansion. However, the accuracy of  MAs for small to intermediate volumes remains unknown and in particular how different MA methods perform with respect to each other. 
Moreover, it is unknown under which conditions MAs give physically meaningful results. In an empirical study
\cite{Schnoerr2014} formulated a set of validity conditions guaranteeing MAs to give physically meaningful approximations to the moments of the CME. We will adopt these validity conditions here. Specifically, whenever the CME  has a stationary solution, we require the MAs to have a single positive and globally attractive fixed point, and their time trajectories to stay non-negative and finite for all times and all initial conditions.
In \cite{Schnoerr2014} it was found that the normal MA fails to satisfy these validity conditions for certain systems and parameter regimes. It was shown that the normal MA can give rise to unphysical behaviour outside of this regime, such as negative mean values or variances, divergent time trajectories, unphysical oscillations and unphysical bistability, thus not allowing for a physical interpretation in these cases. It remains unclear if this is also the case for other moment-closure schemes and how their ranges in parameter space for which they are valid (if they exist) compare to each other. 

In this article, we apply the normal, Poisson, log-normal and CMN-MAs to several chemical reaction systems. We confine our analysis to MAs of second order, since these are the most used in practice. We study their qualitative behaviour in the sense of the validity conditions stated before and compare their quantitative accuracy with exact stochastic simulations \cite{Gillespie1977}.  
It should be stressed that ``validity'' and ``accuracy'' are not unrelated properties, since one can only speak of a method's accuracy when it gives physically valid results. Yet physically meaningful results can be quantitatively highly inaccurate. Therefore, ``validity'' is a necessary but not sufficient condition for ``accuracy''. In this study, we first
use the different MA methods to study stochasticity in a system whose large volume limit is deterministically bistable. Next, we investigate how well the MA methods can capture the influence of noise in a protein-phosphorylation system whose deterministic system shows ultrasensitivity. And finally, we use the MAs to study the role of stochasticity  in a system whose deterministic system is oscillating and which becomes entrained by an external force for a finite time interval.

The derivation of the moment equations from the CME and the subsequent application of moment-closures is conceptually a straightforward task. Practically, however, it becomes extremely cumbersome if more than one species is involved and if one considers higher-order MAs. Suppose for example a system of three species for which we want to compute the fourth-order normal MA equations. Taking symmetries into account, this leads to 34 moment equations which have to be derived from the CME. These will have to be closed, and several fifth-order moments (and potentially higher-order moments) will have to be replaced in terms of lower-order moments. Obviously, this task quickly becomes unfeasible to do manually. Moreover, the numerical analysis of MA equations is not straightforward, and there is no user-friendly software package available allowing non-expert users to derive and analyse MAs. 

To our knowledge, there are three software packages available in the literature for moment-closures: the Matlab toolbox StochDynTools \cite{Hespanha2007} which allows the derivation of MA equations using several different closure schemes for mass-action chemical systems, i.e., those with polynomial propensity functions; the Python package MomentClosure \cite{Gillespie2009} which allows the same but only using the normal moment closure and has the facility to export the MA equations to a Maple file for further analysis; and a Matlab toolbox presented in \cite{Azunre2011} which allows to use normal moment closure to second order for mass-action chemical systems.  For the application of all three packages, the user needs to be familiar with the respective programming language and the numerical analysis is not automated. 

In this article we present the Mathematica package MOCA (moment-closure analysis) which was used for the  presented numerical analysis. MOCA significantly extends the applicability and functionality of the two software packages StochDynTools and MomentClosure \cite{Hespanha2007,Gillespie2009} as well as the software package presented in \cite{Azunre2011}. 
It allows the non-expert user to apply and compare different moment-closure schemes in a graphical user interface (GUI) without any coding necessary. It implements the normal, Poisson, log-normal and CMN-MA and in addition allows the user to define his own novel moment-closure schemes. It extends the applicability to reaction systems with non-polynomial or time-dependent propensity functions. These can either be functions in time or given by discrete time points, for example obtained from experiments. All functions are available either in a GUI or as code version for more experienced users, making the usage of MOCA maximally flexible. MOCA can perform steady state analysis with parameter scans, numerically integrate the MA equations in time and allows to export tables and figures to various commonly used formats.

This paper is structured as follows. Section \ref{sec_methods} introduces the theoretical background for general moment-closure schemes and defines the particular MA methods analysed in this work. The numerical analysis of the various MAs is then presented in Section \ref{sec_numerical_analysis}. Next, we introduce the software package MOCA in Section \ref{sec_moca}. We explain the user input format and demonstrate its capabilities. We finish by summarising our results and concluding in Section \ref{sec_conclusion}.

\section{Background}\label{sec_methods}
\subsection{The chemical master equation and moment-closure approximations}

Consider a chemical reaction system with species $X_i$ ($i=1,\ldots, N$) and $R$ chemical reactions: 
\begin{align} 
  \sum_{i=1}^N  s_{ij} X_i  \xrightarrow{\quad k_j \quad} \sum_{i=1}^N r_{ij} X_i, 
   \quad j = 1, \ldots,  R.
\end{align}
Here, $k_j$ is the rate constant of reaction $j$. We define the elements of the stoichiometric matrix $S$ as 
\begin{align}\label{stoch_mat}
  S_{ij}  =  r_{ij} - s_{ij}.
\end{align}
Under well-mixed and dilute conditions the dynamics of the system is governed by the chemical master equation (CME) \cite{Gillespie2007}:
\begin{align}\label{cme}
  \partial_t P(\mathbf{n},t) 
  & = 
    \sum_{r=1}^R f_r (\mathbf{n} - \mathbf{S}_r) P(\mathbf{n} - \mathbf{S}_r, t) 
    - \sum_{r=1}^R f_r (\mathbf{n}) P(\mathbf{n}, t).
\end{align}
Here, $P(\mathbf{n},t)$ is the joint probability distribution at time $t$,  where $\mathbf{n}=(n_1, \ldots, n_N)$ is the state vector of the system and $n_i$ is the number of molecules of species $X_i$. $\mathbf{S}_r$ is the $r$th column vector of the matrix $S$ and $f_r(\mathbf{n})$ is the propensity function of reaction $r$. For reactions described by the law of mass-action,  the propensity is polynomial and defined as \cite{vanKampen}:
\begin{equation}\label{general_propensity}
f_r(\mathbf{n}) = k_r V \prod_{k=1}^N \frac{n_k!}{(n_k - s_{kj})! V^{s_{kj}}}.
\end{equation}
Here, $V$ denotes the volume of the compartment in which the reaction occurs. 
If in addition $\sum_{i=1}^N  s_{ij} \leq 2$, which basically means that not more than two molecules react which each other in a single reaction (at most a second-order reaction), we call reaction $j$ an ``elementary reaction''. Higher-order reactions do not really occur under conditions found in living cells and although they can often give a useful description of a system, they should really be interpreted as an effective approximate description of a set of elementary reactions, valid only under certain conditions. 

Multiplying \eqref{cme} with $n_i \ldots n_l$ and summing over all $n_i$  $(i=1, \ldots, N)$ leads to the time evolution equation of the moment $\langle n_i \ldots n_l \rangle$:
\begin{align}
  \partial_t \langle n_i \ldots n_l \rangle
  & = 
    \sum_{r=1}^R \langle (n_i+S_{ir})\ldots(n_l+S_{lr}) f_r (\mathbf{n}) \rangle
    - \sum_{r=1}^R \langle n_i \ldots n_l f_r (\mathbf{n}) \rangle.
\end{align}
Here, $\langle \cdot \rangle$ denotes the expectation with respect to $P(\mathbf{n},t)$. Accordingly, the first two moments obey
\begin{align}\label{general_eq_first_moment}
  \partial_t \langle n_i \rangle
  & = 
     \sum_{r=1}^R S_{ir} \langle f_r(\mathbf{n}) \rangle, \\
\label{general_eq_second_moment}
  \partial_t \langle n_i n_j \rangle
  & =
    \sum_{r=1}^R  \big[ S_{jr} \langle n_i f_r(\mathbf{n})  \rangle + S_{ir} \langle  f_r(\mathbf{n}) n_j \rangle
    + S_{ir} S_{jr} \langle  f_r(\mathbf{n})   \rangle \big].
\end{align}
We see that, unless all $f_r(\mathbf{n})$ are a zeroth or first-order polynomial in $\mathbf{n}$, the evolution equation of a certain moment depends on higher order moments, i.e., the equations are not closed. 

The idea underlying the class of moment-closure approximations studied in this work is to express all moments above a certain order $M$ as functions of lower-order moments. The latter is typically done by assuming the distribution of the system to have a particular functional form, for example a normal distribution. This decouples the equations of the moments up to order $M$ from higher-order moments, which then allows one to numerically integrate the moment equations. We refer to such a moment-closure as ``MA of order $M$''. Let

\begin{align}
  y_{i_1,\ldots,i_k}
  &  = 
    \langle n_{i_1} \ldots n_{i_k} \rangle, \\
  z_{i_1,\ldots,i_k} 
  & = 
    \left\{\def\arraystretch{1.2}%
	  \begin{array}{@{}c@{\quad}l@{}}
  	  	\langle (n_{i_1}-y_{i_1}) \ldots (n_{i_k}-y_{i_k}) \rangle & \quad \text{if} \quad k \geq 2, \\
  	 	y_{i_1}  & \quad \text{if} \quad k=1, \\
	  \end{array}\right. \\
\label{def_cum}
  c_{i_1,\ldots,i_k}
  & =
    \partial_{s_{i_1}} \ldots \partial_{s_{i_k}} g(s_1,\ldots, s_N)|_{s_1,\ldots, s_N=0} ,
\end{align}
denote the raw or ``normal'' moments, central moments and cumulants of order $k$ of the system, respectively. 
We call $y_{i_1,\ldots,i_k}$ a ``diagonal moment'' if $i_l = i_m$ for all  $l,m \in \{1,\ldots, k \}$, and a ``mixed moment'' otherwise, and similarly for central moments and cumulants.
In Eq.~\eqref{def_cum} $g(s)$ is the cumulant generating function defined as 
\begin{align}
  g(s_1,\ldots, s_N) = \log \langle \exp (s_1 n_1 + \ldots + s_N n_N) \rangle.
\end{align}
We note that all three types of moments are respectively invariant under permutations of their indices. Therefore, only one representative combination of each permutation class has to be considered. Taking this symmetry into account significantly reduces the number of variables and moment equations. We adopt here the convention that the indices are ordered from small to large, i.e., for a moment $y_{i_1,\ldots,i_k}$ we have $i_1 \leq i_2 \leq \ldots \leq i_k$.
Expressing the moment-closure functions in terms of cumulants rather than raw moments often gives shorter expressions. The equations for the cumulants can then be rearranged to give equations for the raw moments. We consider here the following MA methods: 
\begin{itemize}
\item ``Normal moment-closure" (also called ``cumulant neglect moment-closure" in the literature): all cumulants above order $M$ are set to zero, i.e.,
\begin{align}\label{ma_def_normal}
  c_{i_1, \ldots, i_k} = 0, \quad \text{for} \quad k>M.
\end{align}
\item ``Poisson moment-closure": the cumulants of a one-dimensional Poisson distribution are all equal to the mean value. We assume here the multi-variate distribution to be a product of uni-variate Poisson distributions. Accordingly, for the Poisson MA of order $M$ we set all diagonal cumulants to the corresponding mean and all mixed cumulants to zero, i.e.,
\begin{align}\label{ma_def_poisson_1}
  c_{i_1, \ldots, i_k} 
  & = 
    y_i, \quad \text{for} \quad k>M \quad \text{and} \quad  i_1, \ldots, i_k=i, \quad \text{for some} \quad i \in \{1,\ldots, N \}, \\
\label{ma_def_poisson_2}
  c_{i_1, \ldots, i_k} 
  & = 
    0, \quad  \text{for} \quad k>M \quad \text{and} \quad i_m \neq i_n \quad \text{for some} \quad m,n \in \{1,\ldots, N \}.
\end{align}
\item ``log-normal moment-closure": let $\mathbf{m}$ and $S$ be the mean vector and covariance matrix of a multi-dimensional normal random variable. Then the logarithm of the latter has a multivariate log-normal distribution and its moments can be expressed in terms of $\mathbf{m}$ and $S$  as \cite{Edwin1988}
\begin{align}
  y_{i_1, \ldots, i_k} = \exp \left( \mathbf{v}^T \mathbf{m} + \frac{1}{2} \mathbf{v}^T S \mathbf{v}\right), \quad \text{for} \quad k>M,
\end{align}
where $\mathbf{v} = (g_1, \ldots, g_N)$, where $g_m$ is the number of $i_j$'s having the value $m$. This allows to express $\mathbf{m}$ and $S$  in terms of the first two moments $y_i$ and $y_{i,j}$ which then in turn allows to express higher-order moments in terms of the latter, too.

\item ``CMN moment-closure": all central moments above order $M$ are set to zero:
\begin{align}\label{ma_def_cmn}
  z_{i_1, \ldots, i_k} = 0, \quad \text{for} \quad k>M.
\end{align}

\end{itemize}
Each of the equations can then be used to express the raw moments above order $M$ in terms of lower order moments and thus close the moment equations according to the corresponding MA. We note that the normal MA, Poisson MA and CMN-MA can be equivalent depending on the reaction system (see later for examples of such systems).

The normal moment-closure has been used in the field of biochemical reactions for example in \cite{Verghese2007} and is probably the most commonly used one. It has been considered together with the Poisson and log-normal MA for the one-dimensional stochastic logistic model in \cite{Nasell2003}. The log-normal moment-closure technique has been proposed in \cite {Keeling2000}. In \cite{Singh2006} it has been shown that the assumption of a log-normal distribution is equivalent to a ``derivative matching" closure. The CMN-MA has also been called a ``low dispersion moment-closure" in \cite{Hespanha2008}.

\subsection{Example} 

As an example, consider a reaction system of the Michaelis Menten type:
\begin{align}\label{michaelis_menten}
   \varnothing \xrightarrow{\quad k_1 \quad}S, \quad S+E \xrightarrow{\quad k_2 \quad} SE
  \xrightarrow{\quad k_3 \quad} E + X,
\end{align}
where $E$ is the free enzyme, $S$ is the substrate, $SE$ is the enzyme-substrate complex and $X$ the product. The sum of the numbers of $E$ and $SE$ molecules is constant at all times since each enzyme is either in the free $E$ or complex $SE$ state.
Let $e_0$ denote the total number of enzyme molecules. Assuming mass-action kinetics, the propensity vector is given by 
\begin{align}\label{mm_propensity}
  \mathbf{f}(n_1,n_2)
  & =
    (V k_1, \frac{k_2}{V} n_1n_2, k_3(e_0-n_2)) = (c_1, c_2 n_1n_2, c_3(e_0-n_2)),
\end{align}
where $V$ is the volume of the system and we have defined $c_1 = V k_1, c_2 = k_2/V$ and $c_3=k_3$. Here, $n_1$ and $n_2$ denote the copy number of substrate $S$ and free enzymes $E$, respectively, and we have used the fact that the number of complex molecules $SE$ is $e_0-n_2$ to eliminate the corresponding variable. The stoichiometric matrix is defined in Eq.~\eqref{stoch_mat} and reads for system \eqref{michaelis_menten}
\begin{align}\label{mm_stoch}
  S
  & =
    \begin{pmatrix}
        1  & -1  &  0 \\
        0  & -1  &  1
    \end{pmatrix}.
\end{align}
The corresponding CME is obtained by substituting Eqs.~\eqref{mm_propensity} and \eqref{mm_stoch} in Eq.~\eqref{cme} leading to
\begin{align}
  \partial_ tP(n_1,n_2, t)
  & = 
     c_1 P(n_1-1,n_2,t) +  c_2 (n_1+1) (n_2+1) P(n_1+1,n_2+1, t)  \\
   & \quad
     + c_3  (e_0-n_2+1) P(n_1,n_2-1, t)
    -    (c_1+ c_2 n_1 n_2 + c_3 (e_0-n_2)) P(n_1,n_2, t) .
\end{align}
Multiplying with $n_1, n_2, n_1^2, n_1 n_2$ and $n_2^2$ and summing over all $n_1$ and $n_2$ gives the following equations for the first two moments
\begin{align} \label{ex_eq_first1}
  \partial_t y_1
  & =
    \partial_t \langle n_1 \rangle
  =
    c_1 - c_2 y_{1,2}, \\
  \partial_t y_2
  & =
    \partial_t \langle n_2 \rangle
  =
    -c_2 y_{1,2} +  c_3(e_0-y_2), \\
  \partial_t y_{1,1}
  & =
    \partial_t \langle n_1^2 \rangle
  =
    c_1+ 2 c_1 y_1 + c_2 y_{1,2} - 2 c_2 y_{1,1,2}, \\
  \partial_t y_{1,2}
  & =
    \partial_t \langle n_1 n_2 \rangle
  =
    c_3 e_0 y_1 + c_1 y_2 + (c_2-c_3) y_{1,2} - c_2 y_{1,1,2} - c_2 y_{1,2,2},\\
\label{ex_eq_second1}
  \partial_t y_{2,2}
  & =
    \partial_t \langle n_2^2 \rangle
  =
    c_3 e_0 + (2 c_3 e_0 - c_3) y_2 + c_2 y_{1,2} - 2 c_3 y_{2,2} - 2 c_2 y_{1,2,2}. 
\end{align}
Recall that the moments are invariant under index permutations and thus $y_{2,1} = y_{1,2}$ does not have to be considered explicitly.
We see that the equations of the mean $y_1$ and $y_2$ depend on the second moment $y_{1,2}$. The equation of the latter depends on the third moments $y_{1,1,2}$ and $y_{1,2,2}$ and similarly the equations for $y_{1,1}$ and $y_{2,2}$. It can easily be seen that this applies also to all higher order moments, i.e., the time-evolution equation of a moment of order $k$ depends on moments of order $k+1$. Therefore, the system of equations is not closed and cannot be solved directly. 

Now consider the normal MA which sets all cumulants above a certain order to zero. If we aim at closing the equations to second-order, we have to set the third-order cumulants to zero
\begin{align}
  c_{i,j,k} = 0, \quad \text{for} \quad i,j,k=1,2.
\end{align}
Expressing the cumulants in terms of raw moments, this allows one to find expressions of the third-order moments in terms of first and second-order moments. For $y_{1,1,2}$, for example, this reads
\begin{align}
  y_{1,1,2}
  & = 
   2 y_1 y_{1,2} + y_2 y_{1,1} - 2 y_2 y_1^2,
\end{align}
and similarly for the other third-order moments. Replacing the third-order moments accordingly in Eqs.~\eqref{ex_eq_first1} - \eqref{ex_eq_second1} closes the equations. We give here the resulting equations transformed to central moments:
\begin{align}
  \partial_t z_1
  & =
    c_1 - c_2 (z_{1,2}+ z_1 z_2), \\
  \partial_t z_2
  & =
    -c_2 (z_{1,2}+ z_1 z_2) +  c_3(e_0-z_2), \\
  \partial_t z_{1,1}
  & =
    c_1 + c_2 (z_{1,2}+ z_1 z_2) - 2 c_2 (z_2 z_{1,1} + z_1 z_{1,2}), \\
  \partial_t z_{1,2}
  & =
    c_2 z_2(z_1-z_{1,1}-z_{1,2})-c_2 z_1(z_{1,2}+z_{2,2}) + (c_2-c_3) z_{1,2},\\
  \partial_t z_{2,2}
  & =
    c_3 (e_0-z_2 - 2 z_{2,2}) + c_2 z_2 (z_1 -2 z_{1,2}) + c_2 z_{1,2} - 2 c_2 z_1 z_{2,2}.
\end{align}

\section{Numerical analysis} \label{sec_numerical_analysis}

\subsection{Validity conditions}

We recently formulated validity conditions guaranteeing physically meaningful predictions of MA approximations \cite{Schnoerr2014} and analysed the validity of the normal MA for several example systems. We briefly review these conditions here. For a system for which the CME has a stationary solution, the exact moments of the system converge to a single steady-state in the limit of long times. Therefore, for the MAs to be valid moment approximations, we require convergence to a single steady-state in the limit of long times, too. Moreover, the trajectories should preserve a positive mean and even central moments in the molecule numbers for all times and for all sensible initial conditions. Note that this is also the case for deterministic bistable systems and deterministic oscillatory systems. If the CME converges to a stationary solution, the resulting moments are unique, even if the deterministic rate equations are bistable. Moreover, while single SSA trajectories oscillate for a deterministic oscillatory system, the moments of the CME converge to fixed points because single SSA trajectories get out of phase over time. 

In the following we analyse different MAs with respect to these validity conditions and compare their quantitative accuracy with SSA simulations.

\subsection{A deterministic bistable system}

In \cite{Schnoerr2014} it has been shown that for the deterministic bistable Schl\"ogl model \cite{Schlogl1972}, the normal MA gives physically meaningful results only for an intermediate range of volumes. For smaller volumes it shows negative or diverging trajectories, while it becomes bistable for larger ones. The SSA, in contrast, has a globally attractive positive fixed point and non-negative time trajectories for all volumes.
Here, we study the stochastic properties of the minimal elementary reaction system whose rate equations show bistability \cite{Wilhelm2009}
\begin{align}\label{reactions_bistable}
   & \varnothing \xrightarrow{\quad k_0 \quad} X, \quad   Y \xrightarrow{\quad k_1 \quad} 2X, \\
   &  2X \xrightarrow{\quad k_2 \quad} X+Y, \quad X + Y \xrightarrow{\quad k_3 \quad}Y, \\
   \label{reactions_bistable2}
   & X \xrightarrow{\quad k_4 \quad} \varnothing.
\end{align}
We added the first reaction to the ones given in \cite{Wilhelm2009} to prevent the stochastic system from having an absorbing state for zero molecule numbers. Depending on the parameter values the deterministic rate equations become bistable for this system. All parameter sets used in this section are chosen such that this is the case.
We assume mass-action kinetics here. Since the reactions in Eqs.~\eqref{reactions_bistable}-\eqref{reactions_bistable2} are of order two or lower, their rate functions are polynomials up to order two in the species variables. This means that the time evolution equations of the second-moments depend on the third-order moments, but not on higher-order moments. We thus have to express the third-order moments in terms of first and second-order moments to close the equations to second order. Recall that the second-order normal and CMN-MAs  set all cumulants and central moments above order two to zero, respectively (c.f.~Eqs.~\eqref{ma_def_normal} and \eqref{ma_def_cmn}). Since the third-order cumulant and third-order central moment are identical, the second-order normal MA and CMN-MA are thus equivalent for the reaction system in Eq.~\eqref{reactions_bistable}. This is of course a general result, i.e., for chemical reaction systems with elementary reactions and mass-action kinetics (i.e., reactions up to order two and polynomial propensity functions), the second-order normal MA and second-order CMN-MA are identical.

We thus analyse the normal, Poisson and log-normal MA here. 

\subsubsection{Validity}

\begin{figure}[t]
\includegraphics[scale=0.8]{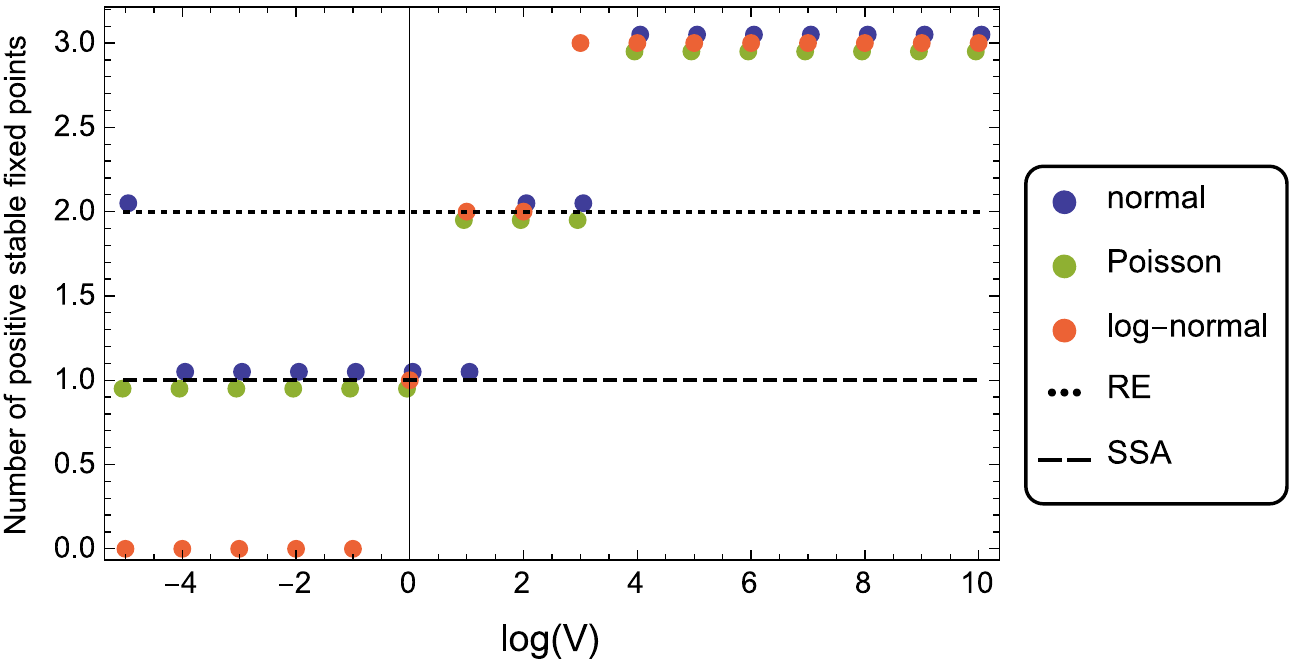}  
\caption{Number of positive stable fixed points as a function of the volume $V$ on log-scale obtained from steady-state analysis for the bistable reaction system in Eqs.~\eqref{reactions_bistable}-\eqref{reactions_bistable2} for the parameters $k_0=1, k_1=1, k_2=5, k_3=0.2$ and $k_4=5$. We shift the points slightly to make coinciding points distinguishable. We find that all three MAs give a physical result of a single positive stable fixed point only on an intermediate range of volumes. The latter is significantly smaller for the log-normal MA than for the normal and Poisson MAs.}
\label{fig_bistable_number_of_fixed_points}
\end{figure}

\begin{figure}[t]
\includegraphics[scale=0.5]{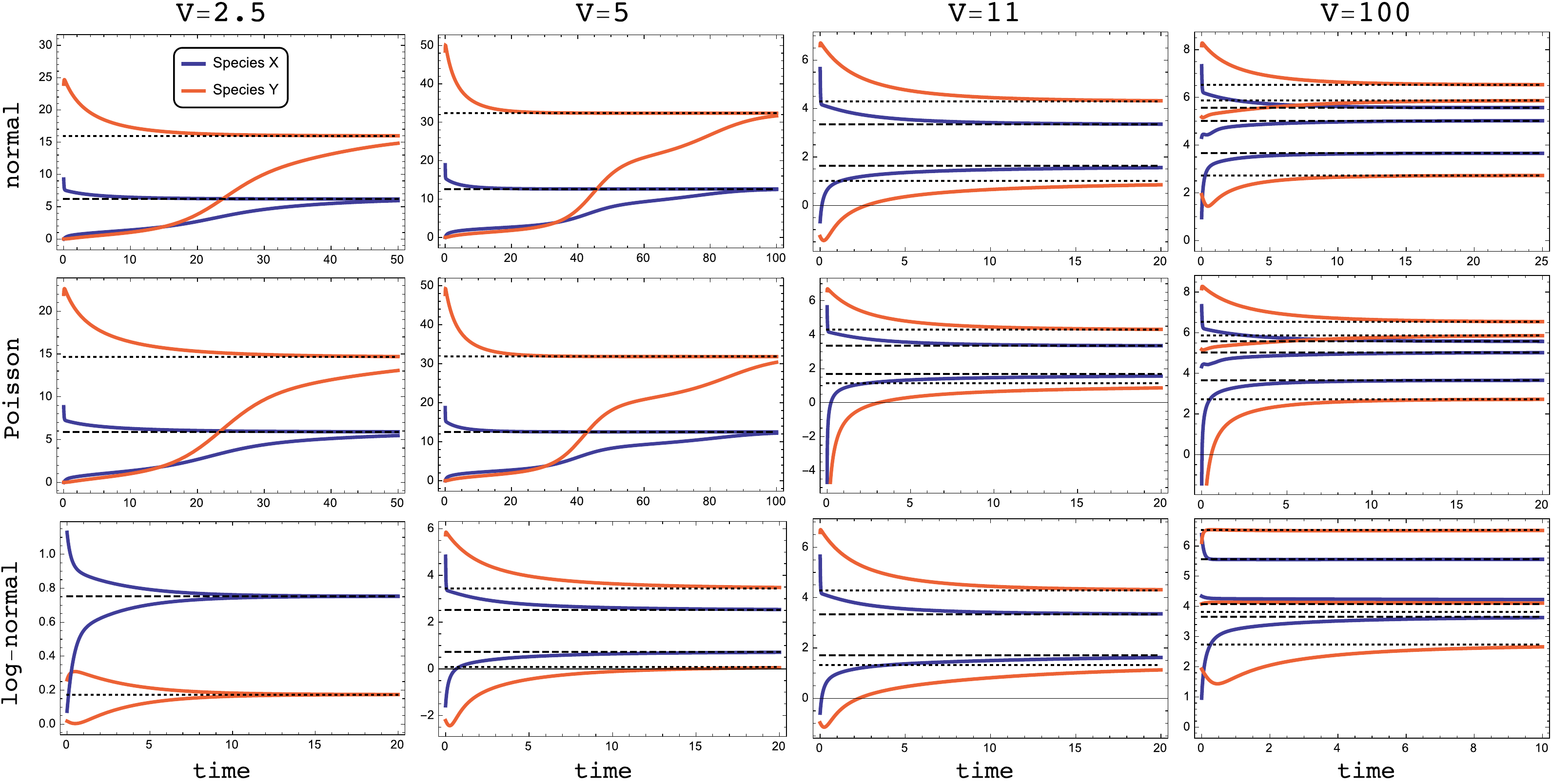}  
\caption{ Time trajectories for the bistable reaction system in Eqs.~\eqref{reactions_bistable}-\eqref{reactions_bistable2} for different volumes $V$ and different initial conditions for the parameters $k_0=1, k_1=1, k_2=5, k_3=0.2$ and $k_4=5$. The dashed and dotted lines indicate the respective positive and stable fixed points of species $X$ and $Y$. Depending on the volume, the MAs have one, two or three positive stable fixed points.
}
\label{fig_bistable_time_course}
\end{figure}

Qualitatively, we find a similar behaviour for the three different MA methods. As for the bistable system analysed in \cite{Schnoerr2014} using the normal MA, we find that there exists an intermediate regime of volumes for the three MAs to be valid, i.e., they have a single globally attractive positive  fixed point, and we find that the moments become bistable (and hence physically meaningless) above this regime. Interestingly, however, we find that when increasing the volume further all three MAs become tristable, i.e., have three positive stable fixed points, see Figure \ref{fig_bistable_number_of_fixed_points}. This means the MAs have more positive stable fixed points than the rate equations here, the latter being bistable independent of the volume, and thus the MAs have no physical interpretation anymore whatsoever. In \cite{Grima2012} it has been shown that for monostable systems, the normal MA becomes equivalent to the rate equations for the means in the limit of large volumes. One can easily show that the result also applies to the Poisson, log-normal and CMN-MA. Here, we find numerically that the tristability remains for volumes up to $10^{10}$, which suggests that the convergence of the MAs to the REs in the limit of large volumes does not hold for deterministic bistable systems. Figure \ref{fig_bistable_time_course} shows the time trajectories for the MAs for different volumes, verifying that the MAs can indeed have one, two or three positive stable fixed points depending on the volume.

The table in Figure 3 lists the endpoints of the validity interval for the MAs for ten different parameter sets on logarithmic scale. The figure below visualises these. We observe that the log-normal MA has a much smaller validity range than the other two MAs.  The normal and Poisson MA most of the time have a similar upper bound while the lower bound is generally smaller for the Poisson MA. We thus find that in terms of validity, the log-normal MA performs significantly worse than the other two MA schemes for the reaction system studied here.

\begin{figure}[t]

\tiny
\begin{center}
  \begin{tabular}{|c|c|c|c|c||c|c|c|c|c|c|c|c|}
  \hline
  \multicolumn{5}{|c} {parameters} &  \multicolumn{2}{|c} {normal} &  \multicolumn{2}{|c} {Poisson} &  \multicolumn{2}{|c|} {log-normal} \\ \hline 
  $~k_0~$ & $~k_1~$ & $~k_2~$ & $~k_3~$ & $~k_4~$ & $\log(V_1)$ & $\log(V_2)$ & $\log(V_1)$ & $\log(V_2)$ & $\log(V_1)$ & $\log(V_2)$   \\ \hline \hline
  $0.5$ & $2$ & $2$ & $0.5$ & $2$ &  $<-11$ & $2.4$ & $<-11$ & $2.4$ & $0.75$ & 1.0 \\ \hline
  $0.5$ & $4$ & $1$ & $0.25$ & $2$ &  $-5.4$ & $-0.59$ & $<-11$ & $-0.11$ & $-1.9$ & 0.58 \\ \hline
  $1$ & $4$ & $1$ & $0.5$ & $2$ &  $<-11$ & $-0.75$ & $-2.5$ & $0.96$ & $0.06$ & 0.49 \\ \hline
  $2$ & $4$ & $2$ & $0.5$ & $4$ &  $-4.7$ & $1.6$ & $<-11$ & $1.7$ & $0.05$ & 0.25 \\ \hline
  $0.25$ & $4$ & $1$ & $1$ & $1$ &  $<-11$ & $2.2$ & $-1.8$ & $2.3$ & $0.73$ & 1.4 \\ \hline
  $1/3$ & $3$ & $3$ & $1/3$ & $3$ &  $-4.9$ & $0.59$ & $<-11$ & $0.57$ & $-1.5$ & 1.3 \\ \hline
  $5$ & $5$ & $1$ & $0.2$ & $5$ &  $-5.8$ & $-0.34$ & $<-11$ & $-0.25$ & $-1.7$ & -0.74 \\ \hline
  $0.2$ & $1$ & $1$ & $0.2$ & $1$ &  $-4.2$ & $1.3$ & $<-11$ & $1.4$ & $-0.13$ & 0.85 \\ \hline
  $1$ & $1$ & $5$ & $0.2$ & $5$ &  $-4.4$ & $1.0$ & $<-11$ & $0.72$ & $-0.38$ & 0.85 \\ \hline
  $0.2$ & $5$ & $5$ & $0.2$ & $5$ &  $-6.0$ & $0.40$ & $<-11$ & $0.40$ & $-3.1$ & 2.0 \\ \hline
  \end{tabular}
  
  \vspace{1cm}
  
  \includegraphics[scale=0.8]{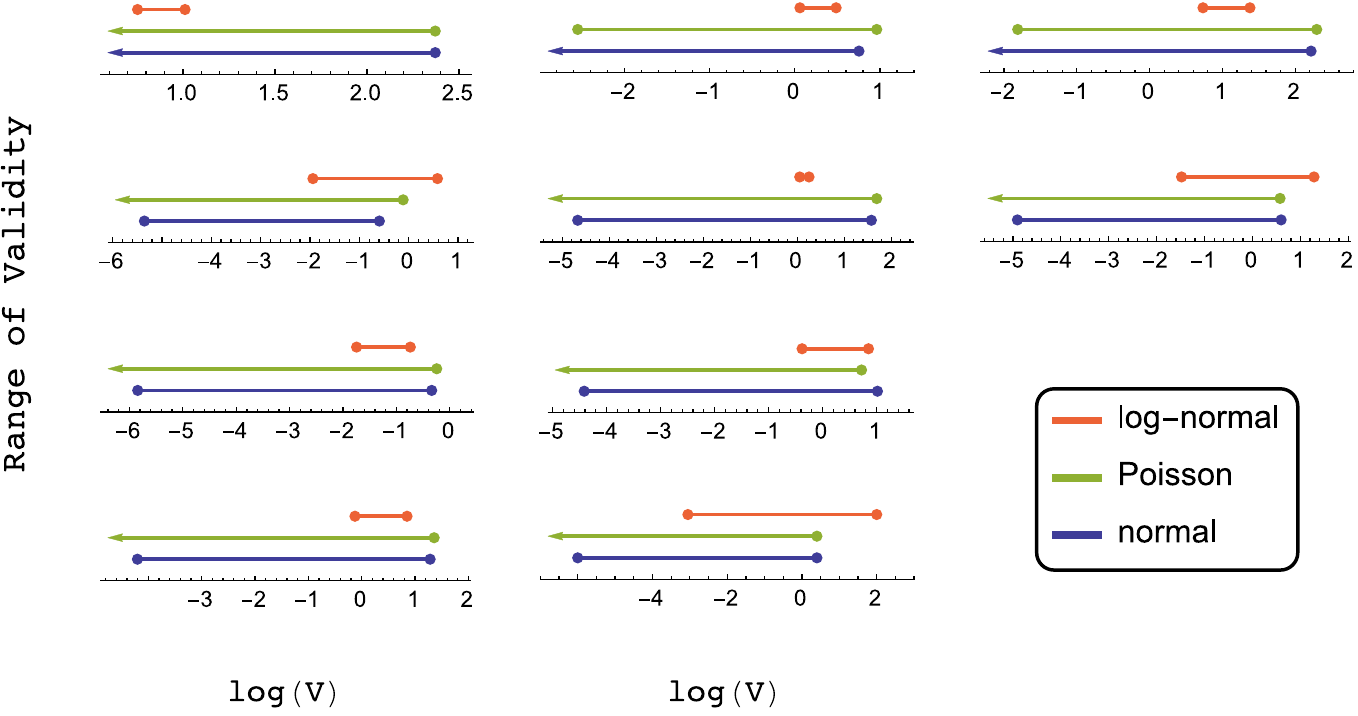} 
  \caption{Top: Range of validity in the volume $V$ on logarithmic scale for different parameter sets for the bistable reaction system in Eqs.~\eqref{reactions_bistable}-\eqref{reactions_bistable2}. $V_1$ and $V_2$ denote the left and right end of the validity interval, respectively. We have only checked for fixed points down to a volume of $e^{-11}$. The term ``$<-11$" thus indicates that the lower boundary of the corresponding validity interval is smaller than $e^{-11}$. Bottom: visualisation of the validity interval on logarithmic scale in the volume for the same ten parameter sets as used in the table. For a lower bound smaller than $e^{-11}$ the lines have an arrow pointing to the left We find that the log-normal MA's range of validity is significantly smaller than that of the normal and Poisson MAs. 
  }
  \end{center}
  
   \label{table_bistable}
\end{figure}

\subsubsection{Accuracy}

\begin{figure}[t]
\includegraphics[scale=0.32]{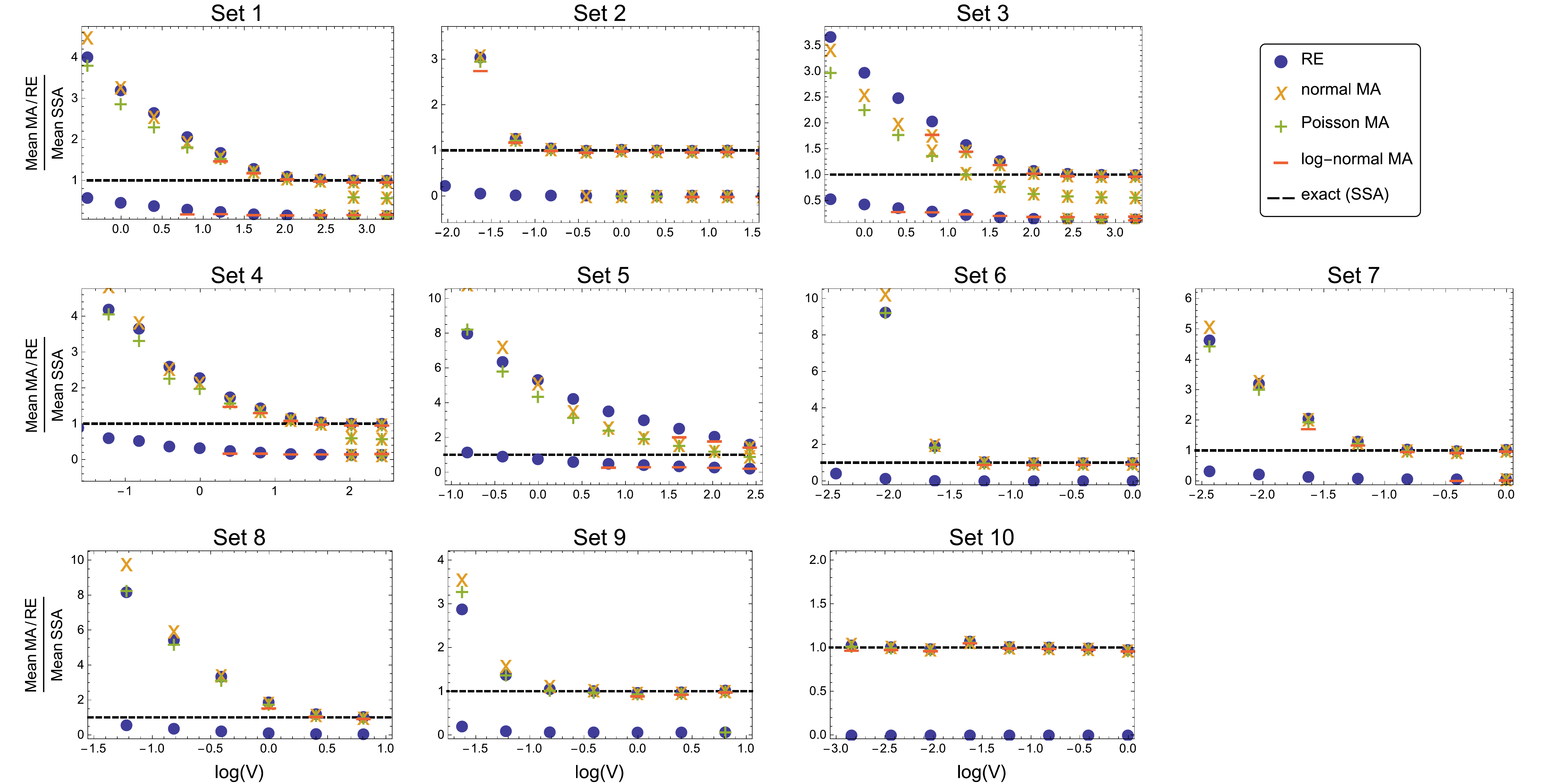}  
\caption{
Mean value of species X in steady state obtained from moment-closures and rate equations as a function of volume $V$ on logarithmic scale for the bistable reaction system in Eqs.~\eqref{reactions_bistable}-\eqref{reactions_bistable2}. The parameter sets are the same as in the table in Figure 3.
The values are divided by the corresponding result obtained from stochastic simulations using the SSA. The horizontal dashed line thus indicates the exact value. For the SSA result $10^4$ samples were simulated for each point.}
\label{fig_bistable_deviation_species_X}
\end{figure}

We next compare the prediction of the different MA schemes and of the rate equations for the mean copy numbers of species $X$ and species $Y$ in steady state with results obtained from exact stochastic simulations using the SSA. The latter have been performed using the software package iNA \cite{Thomas2012}. Figure \ref{fig_bistable_deviation_species_X}  shows the mean values of species $X$ as a function of the volume for the ten parameter sets used in Figure 3. The corresponding figures for species $Y$ look very similar and are not shown here.
The result is divided by the corresponding SSA result. The range of volumes shown corresponds roughly to the validity range of the normal and Poisson MA. We observe here again that the MAs become bistable for larger volumes and that the validity interval of the log-normal MA is significantly smaller than the one of the normal and Poisson MA.

We find that the MAs overestimate the mean copy numbers and that the deviation from the SSA result increases for decreasing volumes. Where two or all three MAs are valid and thus comparable, the accuracy is similar with the log-normal MA being slightly more accurate than the other two and the normal MA being slightly more inaccurate than the Poisson MA. Note, however, that for most parameter sets the log-normal MA's range of validity is significantly smaller than that of the other MAs. 

For large volumes, the MAs have two positive stable fixed points converging to the two positive stable fixed points of the rate equations. The exact result obtained from SSA simulations agrees with the larger of these two fixed points. The third fixed point of the MAs for large volumes seems to always lie between the two of the rate equations. While it lies exactly in the middle for the normal and Poisson MA, it is very close to the lower one for the log-normal MA. We find the same behaviour for all parameter sets. Note though that this can not be seen for all parameter sets in Figure \ref{fig_bistable_deviation_species_X} due to the small plot range.

\subsection{A deterministic ultrasensitive system}

Next, we study an enzyme catalysed protein-phosphorylation system with the reactions
\begin{align}\label{reactions_ultra}
   & P + E_1 \xrightleftharpoons[\quad d_1 \quad]{\quad a_1 \quad } E_1P \xrightarrow{\quad k_1 \quad} P^* + E_1, \\
   \label{reactions_ultra2}
   & P^* + E_2 \xrightleftharpoons[\quad d_2 \quad]{\quad a_2 \quad } E_2P^* \xrightarrow{\quad k_2 \quad} P + E_2.
\end{align}
This system shows ultrasensitivity for certain parameter values \cite{Goldbeter1981}, namely when the enzymes are saturated, i.e., most enzymes are on average in the complex state. Here, $P$ and $P^*$ denote the non-phosphorylated and phosphorylated forms of the protein, respectively, $E_1$ and $E_2$ the phosphorylating and de-phosphorylating enzymes, respectively, and $E_1P$ and $E_2P^*$ the respective protein-enzyme-complexes. In \cite{Goldbeter1981} the authors studied the dependence of the ratio of phosphorylated to non-phosphorylated proteins as a function of $w_1/w_2$ with $w_1=k_1 E_1^t$ and $w_2=k_2 E_2^t$ in a deterministic system,  where $E_1^t$ and $E_2^t$ are the conserved total numbers of the respective enzymes in the system. Assuming a Hill-type response curve, the corresponding Hill coefficient is often used to quantify the steepness of the response. The authors here speak of an ``ultrasensitive response" whenever the response is steeper than a Michaelis-Menten response, i.e., has a Hill coefficient of larger than unity.

We study here the effect of noise on the ultrasensitive response and again compare moment-closure results with SSA simulations. The latter have been performed using the software package iNA \cite{Thomas2012}. First, however, we describe a surprising non-uniqueness of the Poisson and log-normal MA and study the validity of the different MA schemes. As we have explained below Eq.~\eqref{reactions_bistable2}, the second-order normal and second-order  CMM-MA are identical for elementary reaction systems with mass-action kinetics. Since this is the case here, we study the normal, Poisson and log-normal MAs in the following.

\subsubsection{Non-uniqueness for reduced systems}

The studied reaction system in Eqs.~\eqref{reactions_ultra} and \eqref{reactions_ultra2} has six species: $P, P^*, E_1, E_2, E_1P, E_2P^*$, and three conservation laws: the total number of proteins and the total numbers of the respective enzymes, i.e., $P+P^*+E_1P+E_2P^*$, $E_1 + E_1P$ and $E_2+E_2P^*$, are conserved, where we use the symbol for the species also as the corresponding molecule number variable in a slight abuse of notation. The conservation laws allow one to reduce the system to three variables, which is obviously of computational advantage. There are two ways of obtaining the reduced moment-closure equations: arguably, the standard approach would be to start from the reduced CME, compute the reduced moment equations and subsequently apply the moment closure. Alternatively, one may start from the full CME, compute the moment-closure equations and afterwards reduce the equations by taking the conservation laws into account. One may expect, or require, the two approaches for a sensible moment-closure scheme to be equivalent. It is easy to show that this is indeed the case for the normal and CMN moment-closures. However, we find here that this is not the case for the Poisson and log-normal MA. We thus conclude that \emph{the Poisson and log-normal MAs are generally not uniquely defined if one reduces a system according to conservation laws in molecule numbers}, a clear flaw of these methods. The reason for the non-uniqueness of the MA equations is that while the moment-equations depend on \emph{diagonal} higher-order moments if one starts from a reduced CME, no such dependence is found if the MA equations are derived from the full CME. While the normal and CMN-MAs treat diagonal and non-diagonal moments equivalently, the Poisson and log-normal MAs do not do so, thus leading to the issue of non-uniqueness. We explain this in more detail in Appendix A.

One consequence of this non-uniqueness is that certain symmetries of the system are broken. Looking at the reaction system in Eqs.~\eqref{reactions_ultra} and \eqref{reactions_ultra2} one sees that the system is symmetric under exchanging species labels and reaction constants, $P \leftrightarrow P^*$ and $E_1 \leftrightarrow E_2$ and $a_1 \leftrightarrow a_2, d_1 \leftrightarrow d_2$ and $k_1 \leftrightarrow k_2$. This means that for $a_1 = a_2, d_1 = d_2$ and $k_1 = k_2$ the mean values of $P$ and $P^*$, $E_1$ and $E_2$, as well as $E_1P$ and $E_2 P^*$ should be respectively equal. We find that this is indeed the case for the normal and CMN moment-closure, and also for the Poisson and log-normal MAs if one derives the equations starting from the full CME. If one applies the Poisson and log-normal MAs to the reduced CME, however, \emph{they do break the symmetry}.

We conclude that one should be careful when using the Poisson or log-normal MA for systems with conservation laws. In case the MAs are non-unique it is favourable to first derive the MAs before applying the conservation laws. In the following we will study the opposite cases, i.e., if the Poisson and log-normal MA are applied to the reduced CME, which would be normally the standard approach.

\subsubsection{Validity}\label{sec_ultra_validity}

As in \cite{Goldbeter1981} we define $w_1=k_1 E_1^t$ and $w_2=k_2 E_2^t$. The authors in \cite{Goldbeter1981} studied the dependence of the fraction of the protein number in the phosphorylated state as a function of $w_1/w_2$ using deterministic rate equations. The authors call this response ``ultrasensitive" whenever it is steeper than Michaelis-Menten response, meaning a Hill-coefficient larger than one. Here, we would like to study the effect of noise on the response and investigate how different moment-closures perform for this system. To this end, we compute the mean value of the phosphorylated protein $P^*$ in steady state using the different methods of the protein on a grid in $w_1/w_2$ with all the other parameters fixed and fit a Hill function $(w_1/w_2)^{n_H}/(K_d+(w_1/w_2)^{n_{H}})$ to the result, where $K_d$ and $n_H$ are the dissociation constant and the Hill-coefficient, respectively.

\begin{figure}[t]
\includegraphics[scale=0.7]{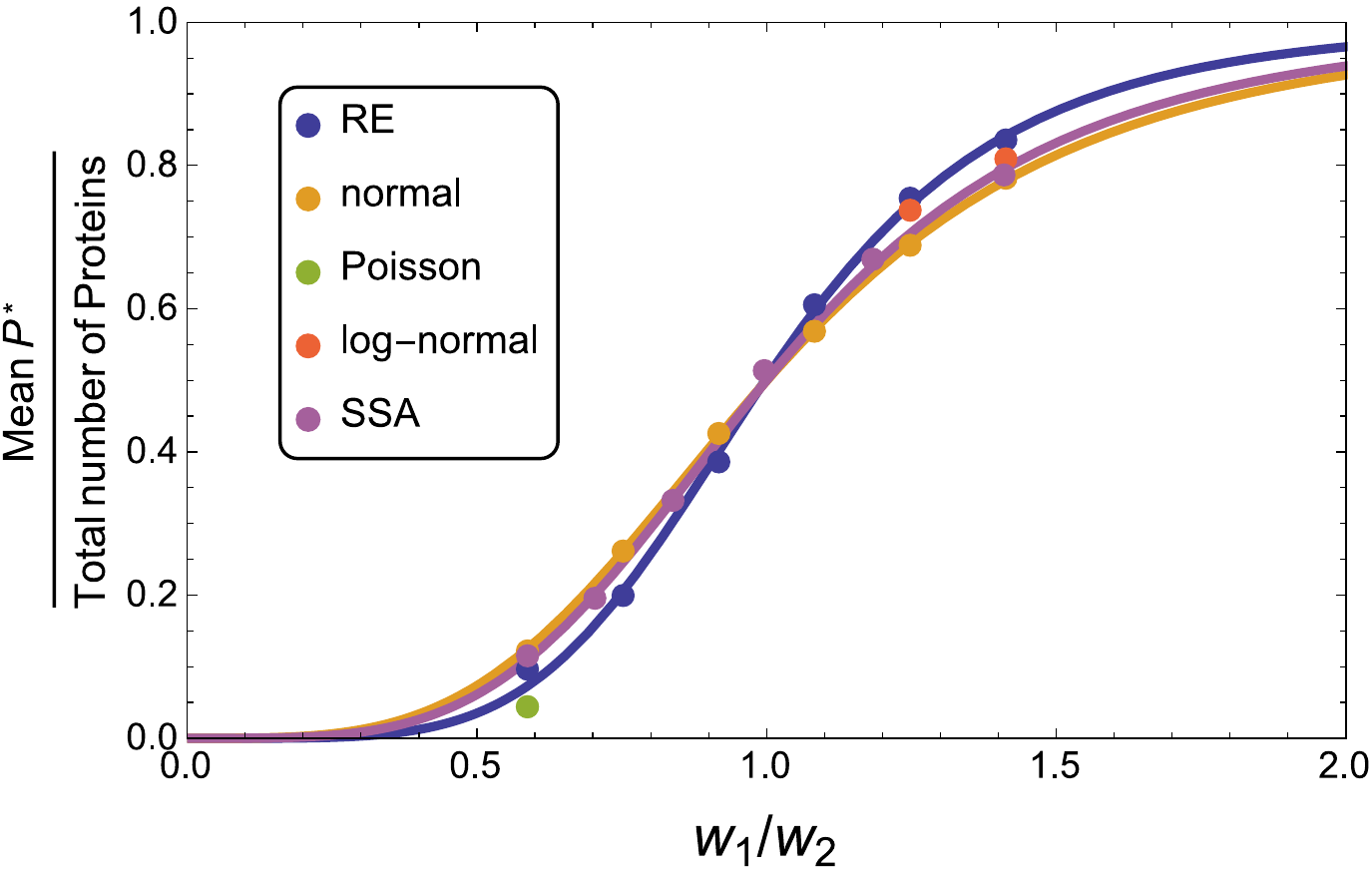}  
\caption{Fraction of mean phosphorylated protein in steady state as a function of $w_1/w_2$ for the protein phosphorylation system in Eqs.~\eqref{reactions_ultra} and \eqref{reactions_ultra2}. The blue and orange curve are Hill-functions fitted to the points of the RE and normal MA, respectively. The Poisson and log-normal MAs have only few positive stable fixed points in the response region making a sensible fit impossible. The used parameters are $a_1=a_2=5, d_1=d_2=1,k_1=k_2=1, V=1, E_1^t=E_2^t=7$ and $P^t=15$, where $E_1^t, E_2^t$ and $P^t$ are the total number of enzyme $E_1$, the total number of enzyme $E_2$ and the total number of proteins in the system, respectively. For the SSA result $10^4$ samples were simulated for each point.}
\label{fig_ultra_example}
\end{figure}

We find that the normal MA and rate equations are valid for all $w_1/w_2$ for all chosen parameter sets, whereas the Poisson and log-normal MA are not for certain parameter regimes, i.e., they do not always have a positive stable fixed point. 
Figure \ref{fig_ultra_example} visualises the fitting procedure for one parameter set. While the rate equations and normal MA are stable on the whole considered response region in $w_1/w_2$, the Poisson and log-normal MAs are unstable for the major part of the region. We obtain only one and two values in the response region, respectively. The Poisson and log-normal MAs thus do not allow a sensible estimate of the response-steepness via a fit of a Hill-function.

\begin{figure}[t]
\includegraphics[scale=0.5]{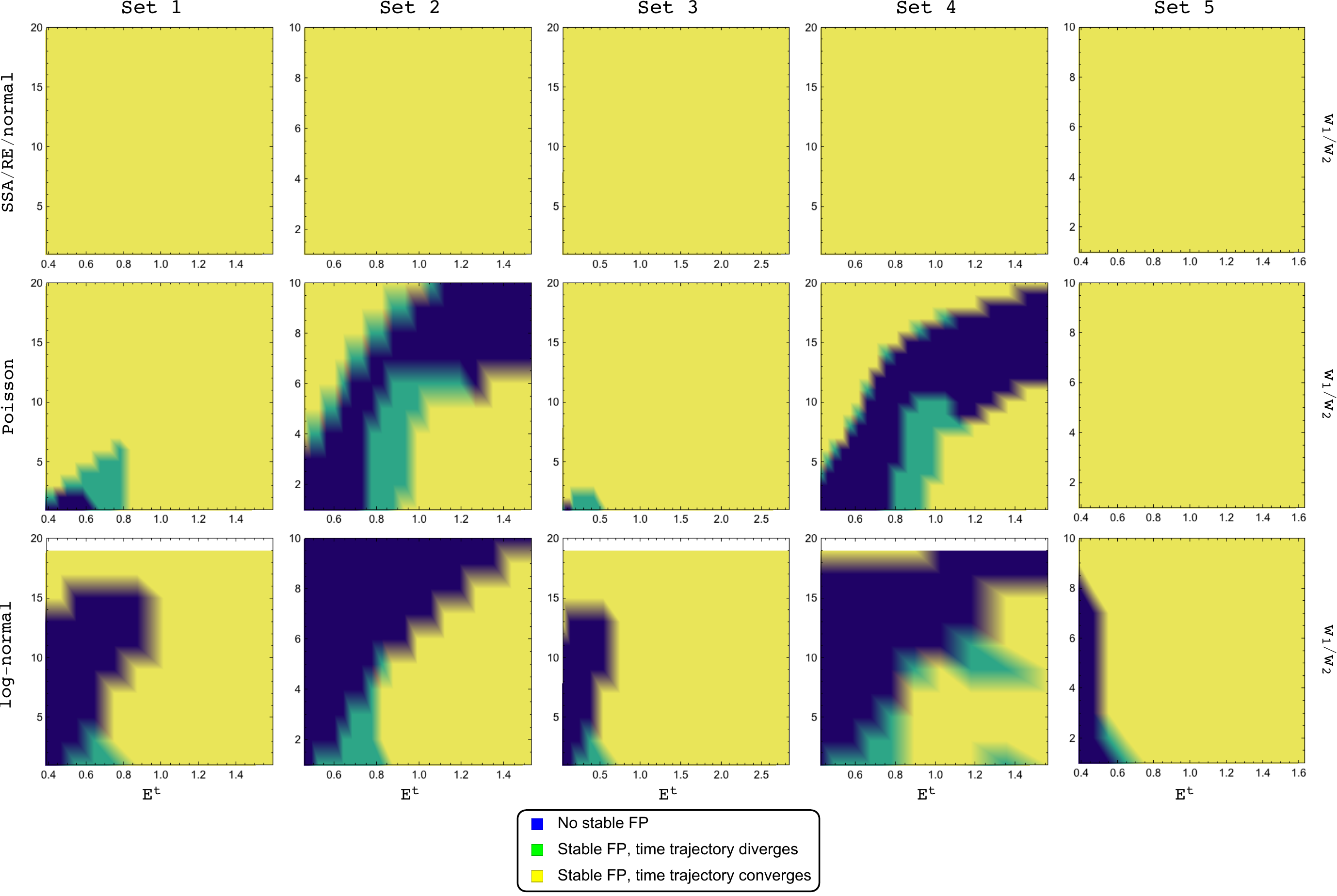}  
\caption{Validity of different MAs as a function of the total enzyme numbers $E_1^t=E_2^t=E^t$ and of $w_1/w_2$ for the protein phosphorylation system in Eqs.~\eqref{reactions_ultra} and \eqref{reactions_ultra2} for five different parameter sets. If we write $(a,d,k,P^t,V)$ with $a_1=a_2=a$, $d_1=d_2=d$ and $k_1=k_2=k$, where $P^t$ is the total protein number and $V$ is the volume, the parameter sets are given by Set~1 =$(1,1,1,25,0.3)$, Set~2 = $(5,1,1,15,1)$, Set~3 = $(5,2,2,25,1)$, Set~4 = $(10,1,1,25,1)$ and Set~5 = $(1,1,1,20,1)$. The blue regions indicate that the methods have no positive stable fixed point. The yellow regions indicate where a positive stable fixed points exists \emph{and} the time trajectories converge with initial condition being the fixed point of the rate equations. The green regions show where the time trajectories diverge despite the existence of a positive stable fixed point, which means that the fixed point is only locally attractive.
}
\label{fig_ultra_validity}
\end{figure}

Figure \ref{fig_ultra_validity} visualises the validity of the rate equations, normal, Poisson and log-normal MAs as a function of the total enzyme number and $w_1/w_2$ for five different parameter sets. The figure indicates where the methods have a positive stable fixed point and where not. In addition, when a positive stable fixed points exists, we solve the time-dependent MAs with the initial condition being the fixed point of the rate equations for the corresponding parameters, and the figure indicates the regions where these diverge despite the existence of a positive stable fixed point. This thus indicates the sensitivity of the different methods to initial conditions. While the rate equations and normal MA are stable and the time trajectories converge everywhere, the Poisson and log-normal MA do so only in subregions of the parameter space. Note that we do not make any statements about \emph{unstable} fixed points here since we investigated the convergence of time-trajectories only for one fixed initial condition. The divergence of the time-trajectories in the green region suggest that there exists an unstable positive fixed point, but the same might be true in some parts of the yellow region despite the convergence of time-trajectories.

In conclusion, we find that the normal MA performs significantly better than the Poisson and log-normal MA for the studied system in terms of validity.

\subsubsection{Accuracy}

\begin{figure}[t]
\includegraphics[scale=0.65]{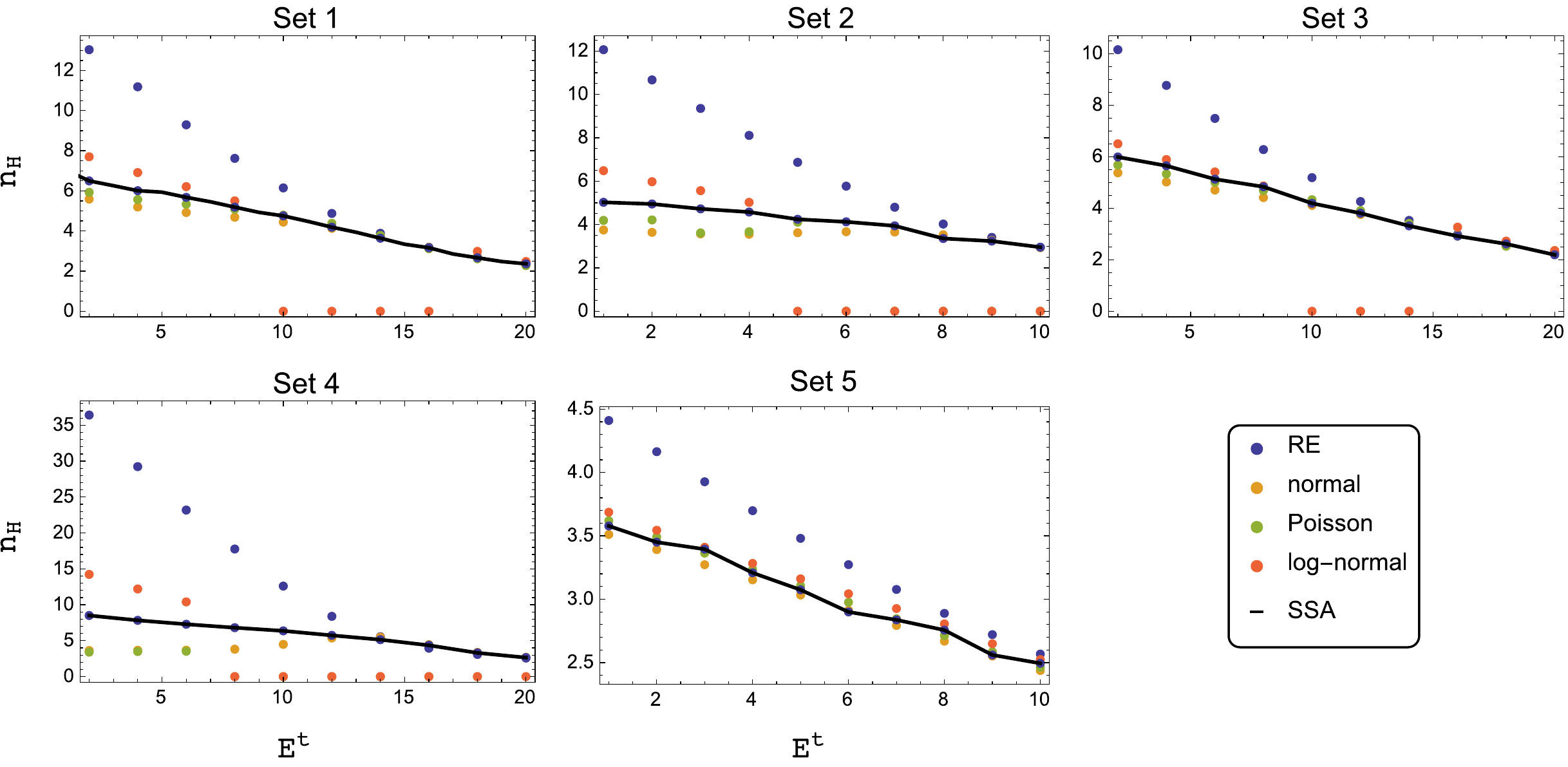}  
\caption{The
Hill coefficient as a function of total enzyme number for the five different parameter sets introduced in Figure \ref{fig_ultra_validity} for the protein phosphorylation system in Eqs.~\eqref{reactions_ultra} and \eqref{reactions_ultra2}. The SSA result is shown as a solid black line.
As explained in the main text, for some parameter values the Poisson and log-normal MA do not allow to estimate a Hill function due to instability. In such cases we set the Hill coefficient to zero. For the SSA result $10^4$ samples were simulated for each point.}
\label{fig_ultra_hill}
\end{figure}

Next, we compare the Hill coefficient obtained from the different methods with the results obtained from SSA simulations as a function of the total enzyme number $E^t$ for the five parameter sets defined in the caption of Figure \ref{fig_ultra_validity}. The SSA simulations were performed using the software package iNA \cite{Thomas2012}. If a method did not allow to estimate a Hill coefficient for some $E^t$ we set the corresponding value to zero. Figure \ref{fig_ultra_hill} illustrates the results. First of all, we find that the rate equations overestimate the Hill coefficient for all $E^t$, with a larger deviation for small $E^t$, which means that the noise in the system significantly reduces the steepness of the response. For small $E^t$ the Hill coefficient estimated from the rate equations becomes up to four times larger then the SSA result (Set 4 in Figure \ref{fig_ultra_hill}). Whenever they allow to estimate a Hill coefficient, the moment-closure approximations are more accurate than the rate equations. While the normal and Poisson MAs underestimate the response, i.e., overestimate the influence of noise, the log-normal overestimates the response. Accuracy wise the three methods perform very similar, the Poisson MA perhaps being slightly more accurate than the other two. However, this slightly higher accuracy of the Poisson MA does not overcome its disadvantage of instability described in Section \ref{sec_ultra_validity}.

\subsection{A deterministic oscillatory system}

Next, we study the Brusselator, a well known deterministic oscillating chemical system given by \cite{Prigogine1968,Lefever1988}
\begin{align}\label{brusselator_reactions}
  2 X + Y \xrightarrow{\quad c_1 \quad} 3X,
  \quad X \xrightarrow{\quad c_2 \quad} Y, 
  \quad \varnothing \xrightleftharpoons[c_4]{\quad c_3 \quad} X.
\end{align}
Depending on the parameter values, the deterministic rate equations show  sustained oscillations, damped oscillations or overdamped oscillations. Single SSA trajectories may show sustained oscillations, while ensemble averages of the SSA always show damped or overdamped oscillations due to the dephasing of independent trajectories. Therefore, a MA can only be interpreted as a valid moment approximation if its trajectories show damped or no oscillations. In \cite{Schnoerr2014} it has been shown that for a parameter set for which the system in Eq.~\eqref{brusselator_reactions} is a deterministic oscillator, the normal MA is valid only for an intermediate range of volumes, with unphysical sustained oscillatory trajectories for larger volumes and either oscillatory or otherwise unphysical trajectories (i.e., divergent or negative trajectories) for smaller volumes. Here, we want to first study the validity of the different MA methods for different parameter sets, and then analyse their behaviour if the system becomes entrained by an external force.  Note that the first reaction in \eqref{brusselator_reactions} is trimolecular, which means that the corresponding propensity function is of third order in the molecule numbers (c.f.~Eq.~\eqref{general_propensity}). The time-evolution equation of the second-order moments thus depend on the third and fourth-order moments (c.f.~Eq.~\eqref{general_eq_second_moment}). Therefore, since the fourth-order central moments and fourth-order cumulants are not identical (in contrast to the third-order ones), the normal and CMN-MAs are not equivalent for the reaction system in \eqref{brusselator_reactions} and we thus analyse all four MAs separately in the following. 

\begin{figure}[t]
\includegraphics[scale=0.5]{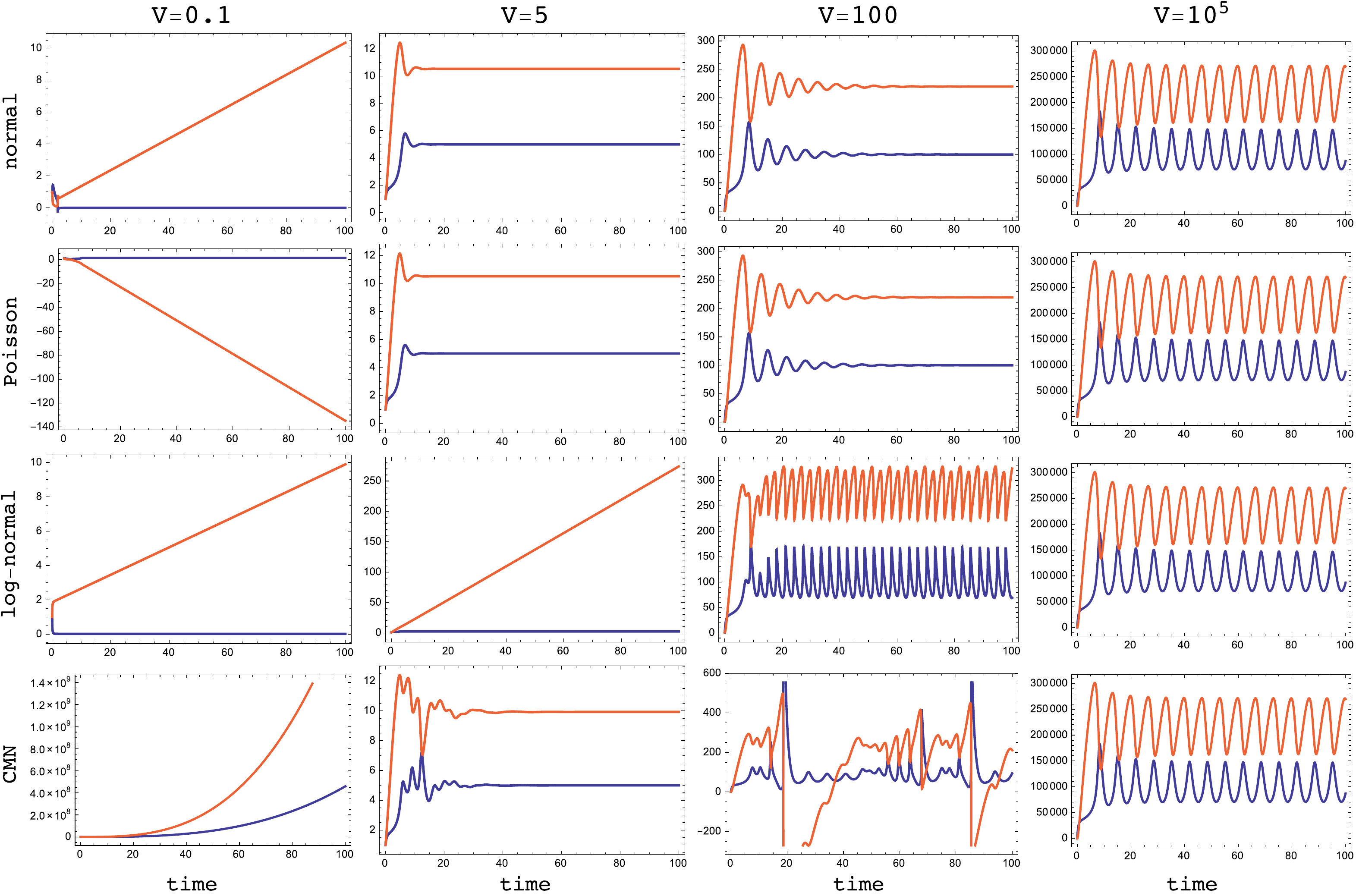}  
\caption{
Time trajectories of the moments of species $X$ (blue line) and $Y$ (orange line) for several volumes for the Brusselator system in Eq.~\eqref{brusselator_reactions} for the parameters $(c_1,c_2,c_3,c_4)=(0.9,2,1,1)$. The blue and red curve denote the mean of species $X$ and species $Y$, respectively. While the normal, Poisson and CMN-MAs
give physically meaningful results, i.e., damped oscillations, for an intermediate range of volumes, the log-normal MA fails to do so for all volumes. To minimise the possibility of numerical effects we computed the shown results using the ODE integration methods ``Adams", ``Backward Differentiation Formula", ``Explicit Runge Kutta", ``Implicit Runge Kutta", 
``Explicit Midpoint" and ``Stiffness Switching" and varied the step sizes over several orders or magnitude, all giving the same results. 
}
\label{brusselator_time_validity}
\end{figure}

\subsubsection{Validity}

\begin{figure}[t]
\includegraphics[scale=0.8]{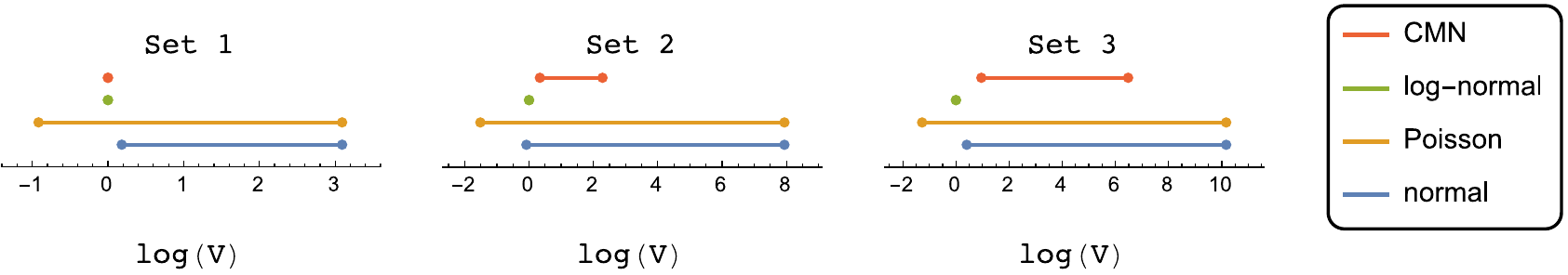}  
\caption{
Range of validity for the Brusselator system in Eq.~\eqref{brusselator_reactions} for three different parameter sets as a function of the volume $V$ in logarithmic scale. The used parameters for $(c_1,c_2,c_3,c_4)$ are Set~1 $=(1,3,0.9,1)$ , Set~2 $=(0.9,2,1,1)$ and Set~3 $=(1,2,1,1.5)$. If the range of validity has length zero we plot a single point at zero. By ``range of validity" we mean the range of volumes for which the MAs give physically meaningful (i.e., non-negative and converging) time-trajectories.
}
\label{brusselator_validity_line_plot}
\end{figure}

We study here the validity of the MAs for three different parameter sets defined in the caption of Figure \ref{brusselator_validity_line_plot}. Similar to the findings in \cite{Schnoerr2014}, we find that all four MAs are only valid on an intermediate regime of volumes. However, unexpectedly, for the log-normal MA we cannot find such a regime. Figure \ref{brusselator_time_validity} shows the time trajectories of the moments for the different MAs for four different volumes for one fixed parameter set. While the normal, Poisson and CMN-MAs diverge for small volumes, are monostable for intermediate volumes and show sustained oscillations for large volumes, the log-normal switches directly from divergent to oscillatory behaviour. We estimated the range of validity for the three different parameter sets for fixed initial conditions of unity for the mean values of both species and zero variance. Figure \ref{brusselator_validity_line_plot} shows the ranges of validity on logarithmic scale in the volume. While the Poisson and normal MA have a finite range of volumes where they lead to physically meaningful results for all parameter sets, the CMN-MA has a vanishing one for one parameter set and the log-normal for all parameter sets.

\subsubsection{System with entrainment}

In systems biology it is frequently of interest to study systems where one or several propensity functions are time-dependent. For example, circadian oscillators are often modelled by a deterministic oscillatory system with an imposed periodic propensity function modelling the influence of an external light input \cite{Westermark2009, Troein2011, Kurosawa2006}. Here, we want to study the performance of the different MA schemes for such a system in the stochastic setting. To this end, we modify the rate constant $c_2$ of the second reaction in Eq.~\eqref{brusselator_reactions} such that it varies over time from $0.5$ to $1.5$ times the chosen mean value in a sinusoidal way, i.e., $c_2(t)=c_2^0 \times(1+\tfrac{1}{2} \sin (\omega t))$ where $c_2^0$ is the fixed mean value of $c_2$ and the frequency $\omega$ of the sine curve is chosen to be the oscillation frequency of the deterministic system. After ten periods, we switch off the time dependence and fix $c_2$ to its mean value. Since we have a time-dependent propensity function here, we cannot use the SSA to simulate the system. We therefore use Extrande, a recently developed exact MC method to sample from the solution of CMEs with time-dependent rate functions \cite{Voliotis2015}. 

\begin{figure}[t]
\includegraphics[scale=0.55]{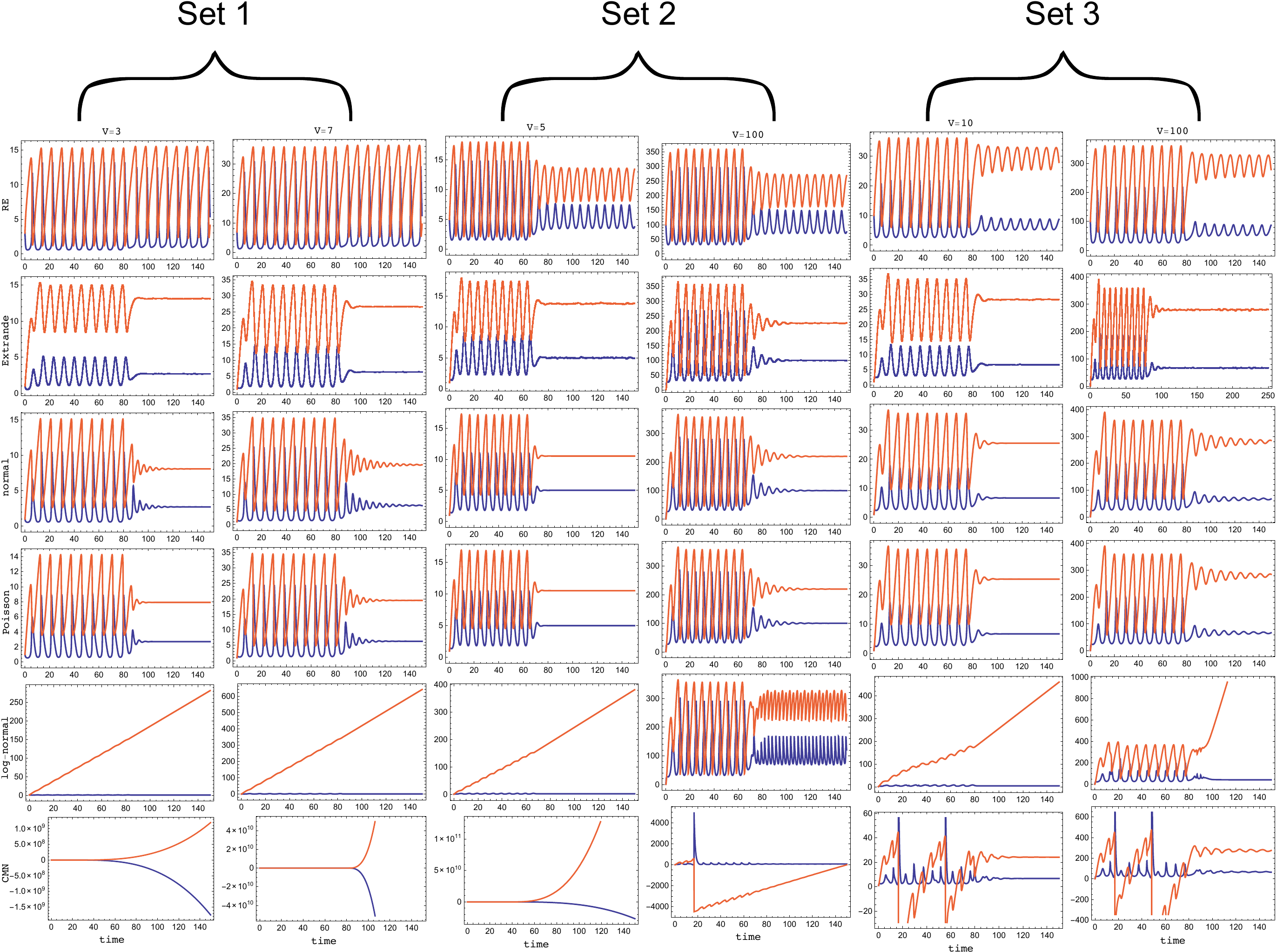}  
\caption{
Time trajectories  for the Brusselator system in Eq.~\eqref{brusselator_reactions} for the three parameter sets defined in the caption of Figure \ref{brusselator_validity_line_plot} with entrainment for two different volumes for each parameter set. The blue and orange lines denote the mean values of species $X$ and $Y$, respectively. The external input gets switched on at time $t=0$ and switched off after ten oscillation periods of the deterministic system (which depends on the given parameter set). While the normal and Poisson MAs
give physically meaningful results (i.e., non-negative and converging time-trajectories) for an intermediate range of volumes, the log-normal and CMN-MAs fails to do so for all volumes. For the Extrande result we simulated $10^5$ samples for Set 1 and $10^4$ samples for Set 2 and Set 3, respectively.}
\label{brusselator_entrainment}
\end{figure}

Figure \ref{brusselator_entrainment} shows the time trajectories for the rate equations, Extrande simulations and the different MA methods.
We find that the rate equations show sustained oscillations after entrainment, while the Extrande results show damped or overdamped oscillations. The normal and Poisson MA behave qualitatively the same way as the Extrande and are thus valid moment approximations for the chosen parameter values. Quantitatively they differ quite significantly from the Extrande result, however. They underestimate the mean values, show oscillations with larger amplitudes during entrainment and a weaker damping after entrainment. Looking at Figure \ref{brusselator_entrainment} one finds that these effects are stronger for the respective smaller volume for each parameter set. The normal and Poisson MA thus \emph{underestimate} the influence of noise here. The log-normal and CMN-MAs fail everywhere to provide a physical result. For the the former this may to be expected, since it also failed to do so in the case without entrainment. Interestingly, however, the CMN-MA is invalid even for parameters for which it is valid in the case without external input. 
Overall, the normal and Poisson MA seem to perform significantly better for this system than the log-normal and CMN-MA.

\section{MOCA}\label{sec_moca}

The Mathematica package MOCA implements the investigated four moment-closure approximations, as well as deterministic rate equations, in a graphical user interface and is freely available in the supplemental material \cite{supplemental}. In contrast to other available moment-closure software packages \cite{Hespanha2007,Gillespie2009, Azunre2011}, MOCA does not only derive the closure equations but also automatically performs numerical analysis of the derived equations, making the methods available to non-expert users. The results are automatically visualised and can be exported to various formats.

\subsection{Applicability}

MOCA extends the applicability over existing moment-closure packages to 
\begin{itemize}
\item  non-polynomial propensity functions 
\item time-dependent propensities functions
\item propensities defined on discrete time points (e.g. measured fluctuating external parameter)
\end{itemize}
Note that while non-polynomial propensities can often give a useful description of a system, they should really be interpreted as an effective approximate description of a set of elementary reactions, valid only under certain conditions \cite{TSG2011}. For these type of propensities the software applies a Taylor expansion of the propensity around the mean value to a specified order as proposed in \cite{Ale2013}. These different features make MOCA applicable to virtually any reaction system with arbitrary propensity functions. 

In addition to the different moment-closure methods described above, MOCA allows the user to define his own moment-closure method, providing an easy way to develop novel moment-closure schemes.

\begin{figure}[t]
\includegraphics[scale=0.8]{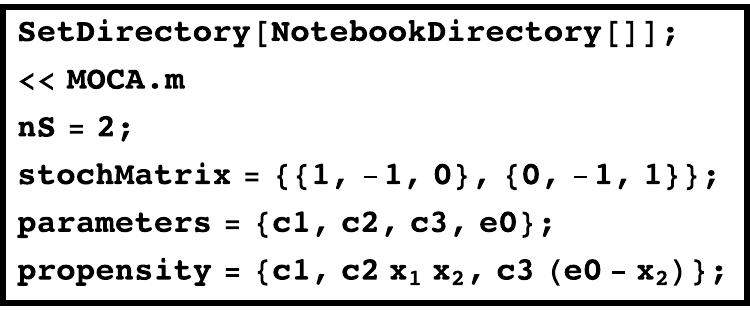}  
\caption{MOCA input for time-independent propensity functions for the example system in \eqref{michaelis_menten}.
The first two lines do not need to be modified. They set the directory of the file and load the package MOCA.m. The following lines define the number of species, stoichiometric matrix,  parameters and propensity functions of the system, respectively. Note that we have absorbed the dependence of the rate functions on the volume $V$ and the rate constants $k_i$ into the parameters $c_i$ as defined below Eq.~\eqref{mm_propensity}.}
\label{input1}
\end{figure}

\subsection{User input}

To use the package, the file MOCA.m needs to be placed in the same folder as the Mathematica notebook that will be used for the analysis. Figure \ref{input1} shows an example input for the corresponding notebook for the reaction system defined in \eqref{michaelis_menten}. The first two lines, which set the path and load the package, respectively, have to be executed without any modification. Next, the number of species and the stoichiometric matrix have to be specified and assigned to the variables \textbf{nS} and \textbf{stochMatrix}, respectively, as depicted in the third and fourth line in Figure \ref{input1}.  The propensity vector and stoichiometric matrix are given in Eqs.~\eqref{mm_propensity} and \eqref{mm_stoch}, respectively. The number of species \textbf{nS} has to be equal to the number of rows of  \textbf{stochMatrix}. Next, the parameter vector \textbf{parameters} and the propensity vector called \textbf{propensity} need to be specified, as done in the fifth and sixth input lines in Figure \ref{input1}.

The species variables have to be denoted by an ``$x$" with the species index as a subscript. All terms in the propensity function that are not species variables or numerical values have to be listed as parameter in \textbf{parameters}. This is all the input needed if dealing with time-independent propensity functions and when using the GUI. Note that the propensities do not need to be of mass-action, i.e., polynomial type, they can have any analytical form. 

For using the coding version of MOCA, deterministic rate equations and time-dependent propensity functions, as well as for the definition of moment-closures, see the corresponding tutorial files in the supplemental material \cite{supplemental}.

\begin{figure}[t]
\includegraphics[scale=0.24]{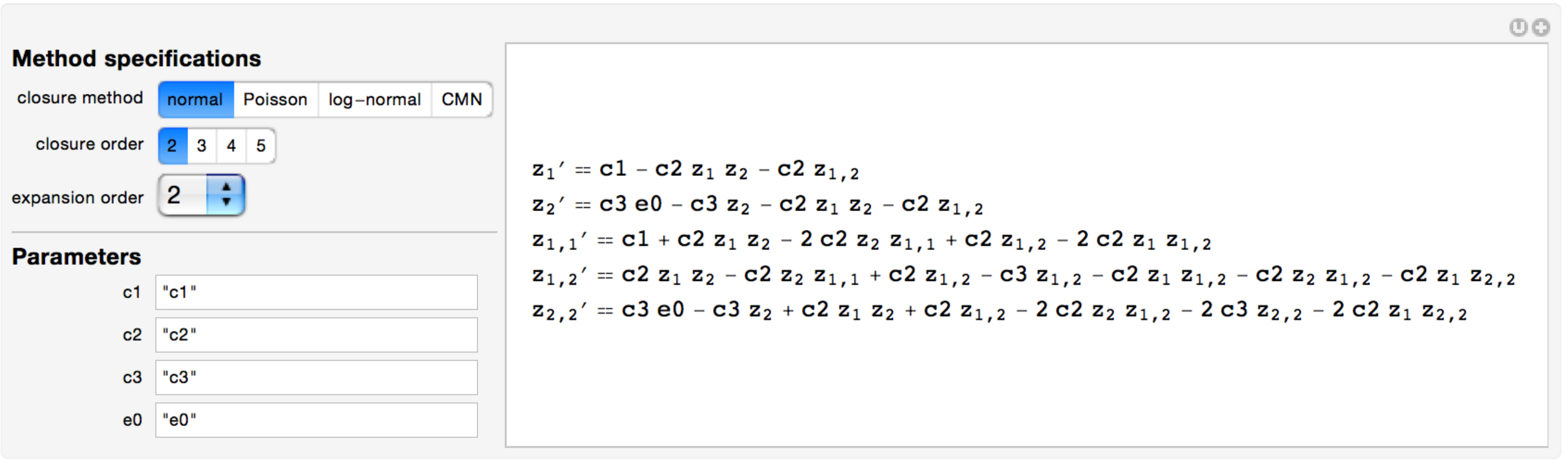}  
\caption{GUI for deriving MA equations with MOCA for the reaction system in \eqref{michaelis_menten}. After defining the system as in Figure \ref{input1} the command 
\textbf{SteadyState} has to be evaluated in the notebook for the GUI to appear. The user can choose the closure method, closure order, expansion order and specify parameter values. For changes to apply the user needs to press the little  ``update button" in the top right corner.}
\label{analysis_mm_equations}
\end{figure}

\subsection{Analysis - the graphical user interface}

There are four functions available to be used within a GUI. They simply need to be typed into the notebook  and evaluated to open the corresponding GUI:
\begin{itemize}
\item   \textbf{DeriveEquations}: derives the MA equations for central moments for general parameters and allows to assign numerical values to the parameters. 
\item \textbf{SteadyState}: numerically searches for positive and stable fixed points of the MA equations.
\item \textbf{SteadyStateVaryParameter}: same as \textbf{SteadyState} but with one parameter varied over a grid specified by the user. The resulting table can be exported into a ``CSV" (``Comma-separated values'') file. 
\item \textbf{TimeTrajectory}: solves MA equations numerically in time for numerical parameter values and plots the result. The result can be exported as a figure to various formats or evaluated on a grid in time and stored in a ``CSV" file.
\end{itemize}
Figure \ref{analysis_mm_equations} shows the GUI that appears after typing and evaluating \textbf{DeriveEquations}. The user can interactively choose a moment-closure method, the closure order as well as the expansion order. By ``expansion order" we mean the expansion of the propensity functions around the mean value as proposed in \cite{Ale2013}. This is only necessary for non-polynomial rate functions. For exclusively polynomial  rate functions, the expansion does not make a difference as long as its order is equal to or higher than the maximum order of the propensity polynomials. Finally, it is possible to assign numerical values to the parameters. The equations only become updated when the small ``update bottom" in the top right corner is clicked. This is also true for the functions described in the following, i.e., changes in the input are only applied after clicking the ``update bottom".

\begin{figure}[t]
\includegraphics[scale=0.25]{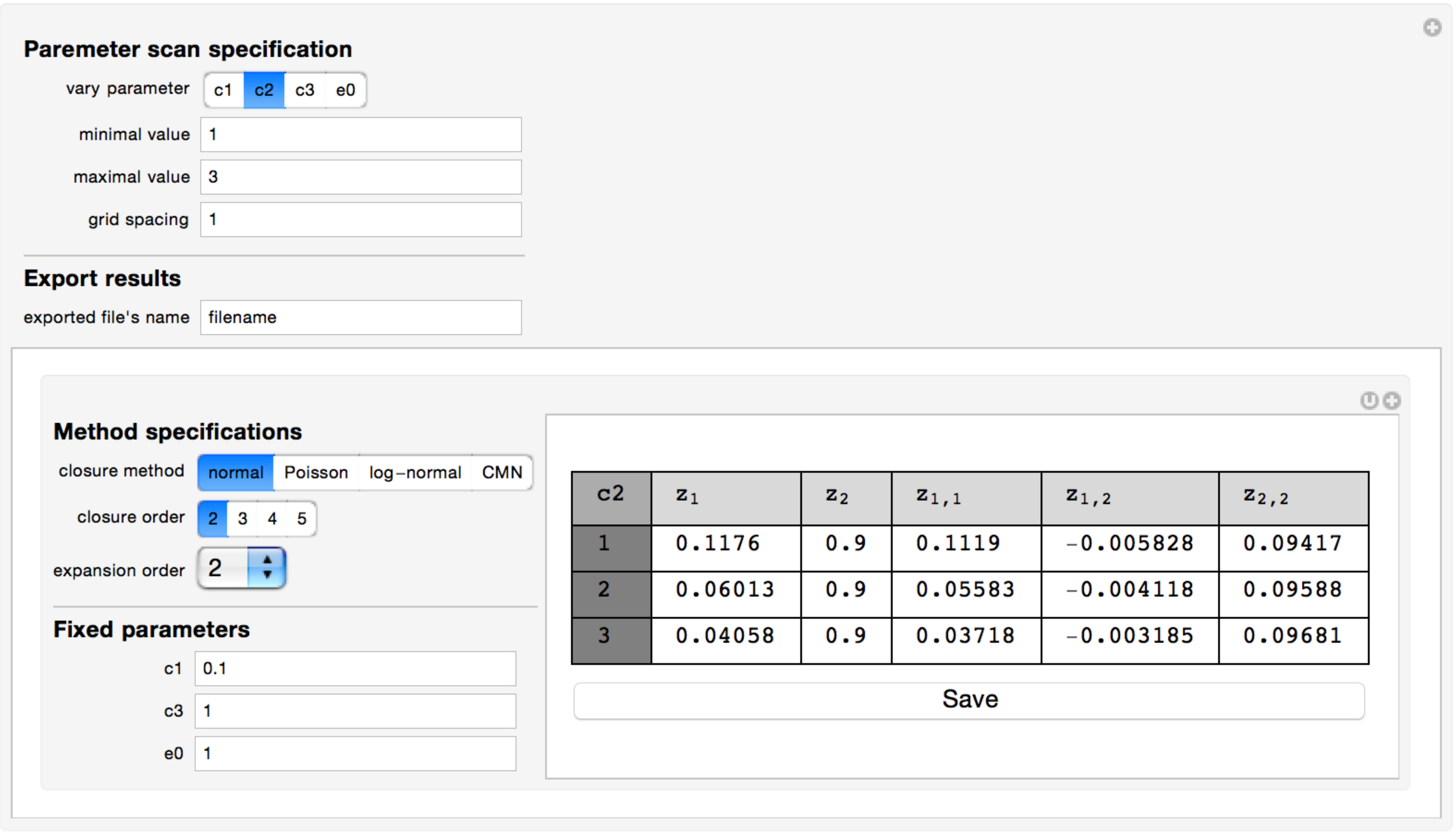}  
\caption{GUI corresponding to the command \textbf{SteadyStateVaryParameter}  in MOCA for the reaction system in \eqref{michaelis_menten}. The table shows positive stable fixed points obtained by varying one parameter over a specified grid.}
\label{analysis_mm_ss_grid}
\end{figure}

The function \textbf{SteadyState} allows to numerically compute positive stable fixed points of the MA equations. It has the same input parameters as the function \textbf{DeriveEquations} described before, with the difference that the parameters have an initial numerical value. For some parameter values, the method cannot find a positive and stable fixed point. However, this does not necessarily mean that the numerical algorithm fails. In \cite{Schnoerr2014} it has recently been shown that MA equations can indeed have no positive and stable fixed point for certain bimolecular reaction system (even though the SSA and rate equations do have positive stable fixed points). The authors also showed that MAs can have more than one positive stable fixed point, in which case \textbf{SteadyState} function may give more than one result.

The function \textbf{SteadyStateVaryParameter} also searches for positive stable fixed points but varies a user specified parameter over a user specified grid. The corresponding GUI is shown in Figure \ref{analysis_mm_ss_grid}. The resulting table can be exported to a text file in ``CSV" format to the same folder where the notebook is located.

\begin{figure}[t]
\includegraphics[scale=0.25]{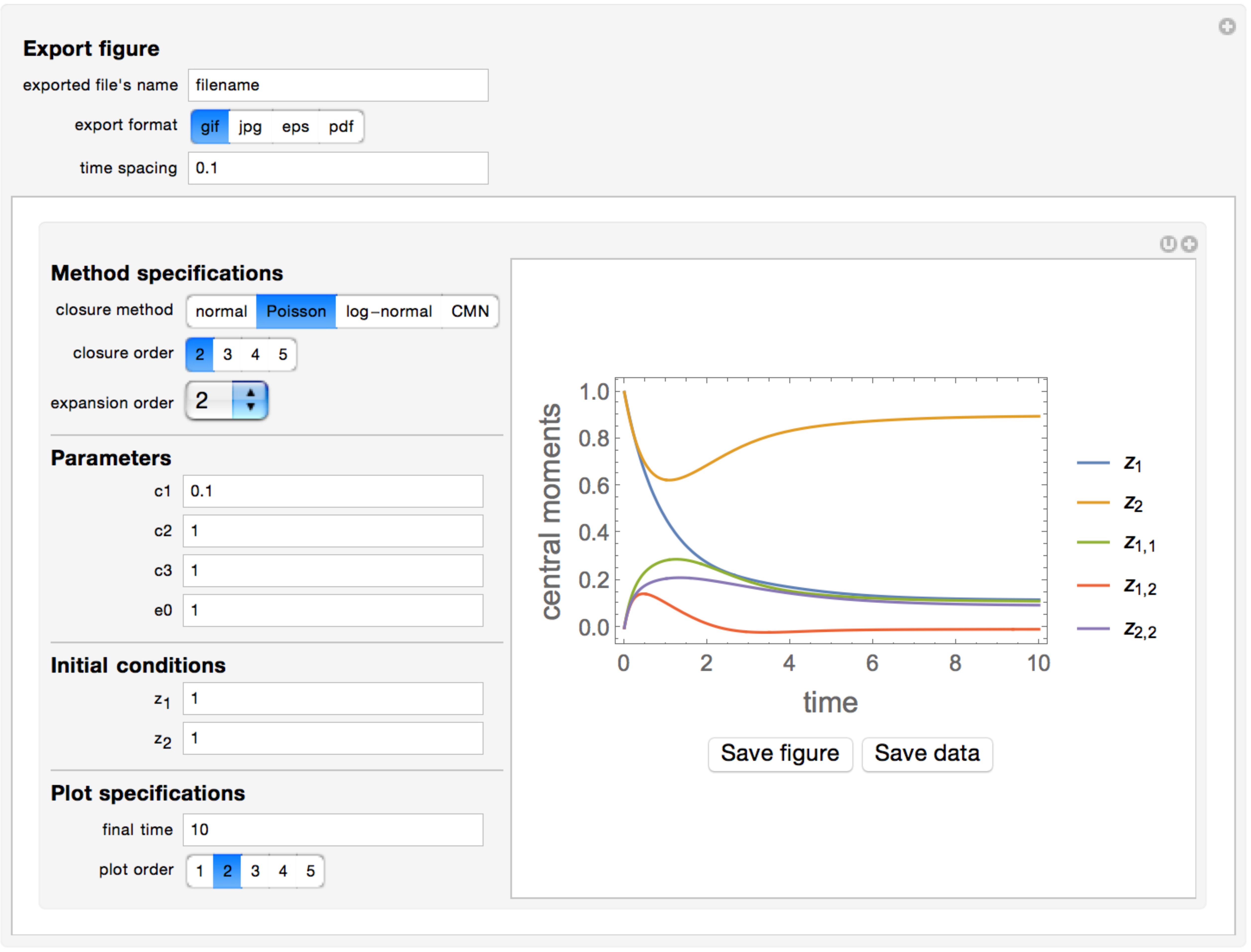}  
\caption{ GUI for solving and visualising the MA equations numerically in time using the \textbf{TimeTrajectory}  command of MOCA. In addition to the method specifications, the user can specify initial conditions for the mean values, the final time point as well as up to which order moments should be plotted. The result can be exported as a figure or into a ``CSV" file evaluated on a time grid.}
\label{analysis_mm_time}
\end{figure}

The final function \textbf{TimeTrajectory} solves the MA equations numerically in time and plots the result. Figure \ref{analysis_mm_time} shows the corresponding GUI. In addition to method specifications and values for parameters, the user can specify initial conditions for the mean values of the species (higher order central moments are set to zero initially, i.e., deterministic initial conditions), the final time point and the plot order specifying up to which order moments should be plotted.  The result can either be exported as a figure to various formats or into a ``CSV" text file where the solution is evaluated on a time grid with user specified time spacing $dt$.

\subsection{Coding commands}

The GUI commands described above are also available as Mathematica functions allowing more experienced Mathematica users a more flexible application of the methods. See the example files in the supplemental material \cite{supplemental} for details on how to use these functions.

\section{Summary and Conclusion}\label{sec_conclusion}

In this paper, we compared the second-order normal, Poisson, log-normal and CMN-MAs for several reaction systems with respect to their qualitative behaviour (if they give physically meaningful results) and their quantitative accuracy (how well they approximate results obtained from exact stochastic simulations) 
 whenever they give physically meaningful results.
While we found no significant difference in quantitative accuracy between the four MAs, the ranges in parameter space for which the MAs gave physically meaningful results were significantly larger for the normal MA suggesting that the normal MA is favourable over the other studied MAs. We emphasise that the presented results are exclusively based on numerical analysis and although we confirmed the results for a wide range of parameter sets and several example systems, we cannot expect them to hold in general for all parameter sets or chemical reaction systems. In \cite{Hespanha2011} for example, it has been found for a single parameter set for one chemical reaction system that the log-normal MA is significantly more accurate than the normal MA. However, for non-linear systems, our results suggest that the MAs give physically meaningful results only above a certain critical volume if the system is deterministically monostable, and only for intermediate volumes if the system is not deterministically  monostable.

By ``physically meaningful" we mean the validity conditions proposed in \cite{Schnoerr2014} which are: (i) the mean values and even central moments of a system should stay non-negative and finite for all times and they should converge to a steady state whenever the CME has a steady state solution, (ii) the moments are unique in the sense that the same steady-state moments can be reached from all initial conditions, and (iii) the moments do not exhibit sustained oscillations in the limit of long times (unless there is an external time-dependent input). In \cite{Schnoerr2014} it has been found that the normal MA does not satisfy (i) for small volumes for several non-linear reaction systems, and that it does not satisfy conditions (ii) and (iii) for large volumes for deterministic bistable and oscillatory chemical systems, respectively. 

Here, we performed a similar analysis for four different MA methods. We first studied a deterministically bistable system, i.e., a system whose rate equations have two positive stable fixed points. Interestingly, we find that the MAs have \emph{three positive stable fixed points} for large volumes, thus allowing no physical interpretation. Surprisingly, we found that, for an enzyme-catalised reaction,  \emph{the Poisson and log-normal MAs were not uniquely defined}. Our analysis suggests that this may indeed be generally the case for systems with conservation laws, a flaw not shared by the other two MAs. Finally, we studied a deterministically oscillatory system with and without an external periodic input. In both cases we found that the normal and Poisson MAs are valid only for an intermediate range of volumes, becoming unstable for smaller volumes and undergoing unphysical sustained oscillations for larger volumes. Curiously, the CMN-MA behaves like this only for some of the studied parameter sets, and the log-normal for none of these, i.e., \emph{there is no range of volumes where the latter two MAs give physically meaningful results}.

In conclusion, our results taken together do not favour one MA over the others in terms of accuracy, but suggest that the normal MA is favourable over the other MAs, in the sense that the range of parameter space where it gives physically meaningful results is considerably larger than that of the other MAs. 

Finally, we presented the software package MOCA which was used for the numerical analysis of the various MAs.  MOCA allows one to derive and analyse moment-closure approximations for systems with polynomial, non-polynomial as well as time-dependent propensities. MOCA implements the ``normal" or ``cumulant-neglect", the ``Poisson", the ``log-normal" and the ``CMN" closures as well as user-defined moment-closure schemes and automatises the numerical analysis. It allows non-expert users to apply moment-closure methods in a user-friendly graphical user interface.

\section*{Acknowledgments}
G.S. acknowledges support from the European Research Council under grant MLCS 306999.

\begin{appendix}

\section{Non-uniqueness for chemical systems with conservation laws}

Here, we investigate in detail, the non-uniqueness of the Poisson and log-normal MAs for systems with conservation laws. To this end, we consider the simple reversible reaction system
\begin{align}
   A + B  \xrightleftharpoons[\quad k_2 \quad]{\quad k_1 \quad } C.
\end{align}
We now compute the MA equations by applying the conservation laws of the system once \emph{after}, and once \emph{before} closing the equations.

\subsection{Closing the equations first}

This approach involves obtaining the moment equations from the CME and subsequently imposing the conservation laws on the resulting moment equations. The stoichiometric matrix $S$ and propensity functions $f_1$ and $f_2$ of the two elementary reactions for this system read (c.f.~Eq.~\eqref{cme})
\begin{align}\label{app_stoch_prop}
  & S
  =
    \begin{pmatrix}
        -1  & 1  \\
        -1  & 1  \\
        1 & -1
    \end{pmatrix}, \\
  & f_1 (n_1,n_2,n_3)
   =
    \frac{k_1}{V}n_1 n_2, \\
  & f_2 (n_1,n_2,n_3)
  =
    k_2 n_3, 
\end{align}
where $n_1, n_2$ and $n_3$ denote the copy numbers of species $A,B$ and $C$, respectively. 
The corresponding time-evolution equations for the first and second-order moments can be obtained from Eqs.~\eqref{general_eq_first_moment} and \eqref{general_eq_second_moment}. For $y_1 = \langle n_1 \rangle$ and $y_{1,1} = \langle n_1^2 \rangle$, for example, they read
\begin{align}
  \partial_t y_1
  & =
    -\frac{k_1}{V} y_{1,2} + k_2y_{3}, \\
  \partial_t y_{1,1}
  & =
    -2 \frac{k_1}{V} y_{1,1,2} + 2 k_2 y_{1,3} + \frac{k_1}{V} y_{1,2} + k_2 y_{3}.
\end{align}
Note that due to the term $n_1 n_2$ in $f_1$, the equation for $y_{1,1}$ depends on the third-order moment $y_{1,1,2}$, but not on any \emph{diagonal} third-order  moment, i.e., not on $y_{1,1,1}, y_{2,2,2}$ or $ y_{3,3,3}$. The same is of course true for the  equations of the other second-order moments: they do not depend on a diagonal third-order  moment. 
This means that the second-order normal and Poisson MAs are equivalent, since they differ only in their expressions for diagonal moments (c.f.~Eqs.~\eqref{ma_def_normal}-\eqref{ma_def_poisson_2}). The corresponding second-order normal and Poisson MAs for $y_1$ and $y_{1,1}$ are obtained by setting the corresponding third-order cumulant $c_{1,1,2}$ to zero which leads to $y_{1,1,2} = 2 y_1 y_{1,2} + y_2 y_{1,1} - 2y_1^2 y_2$ and thus gives
\begin{align}
  \partial_t y_1
  & =
    -\frac{k_1}{V} y_{1,2} + k_2y_{3} \\
  \partial_t y_{1,1}
  & =
     -4 \frac{k_1}{V} y_1 y_{1,2} - 2 \frac{k_1}{V} y_2 y_{1,1} + 4 \frac{k_1}{V}y_1^2 y_2 + 2 k_2 y_{1,3} + \frac{k_1}{V} y_{1,2} + k_2 y_{3},
\end{align}
and similarly for the other first and second-order moments.
Note that the system has two conservation laws
\begin{align}\label{app_cons_1}
   & n_1 + n_3 = A^t = const. , \\
\label{app_cons_2}
   & n_2 + n_3 = B^t = const.
\end{align}
To simplify the following equations, let us assume $A^t=B^t$, which implies $n_1=n_2$.
The system of moment equations of three variables can thus be reduced to a system with only one variable, since all moments of first and second order can be expressed in terms of $y_{1}$ and $y_{1,1}$ using Eqs.~(\ref{app_cons_1}) and (\ref{app_cons_2}). For example, we have $y_3= \langle n_3 \rangle =  \langle A^t - n_1 \rangle =A^t - y_1$ and $y_{1,2} = \langle n_1 n_2  \rangle =\langle n_1 n_1  \rangle = y_{1,1}$ and similarly for the other first and second-order moments. The resulting equations for $y_1$ and $y_{1,1}$ are thus closed and read
\begin{align}\label{app_ma_first_1}
  \partial_t y_1
  & =
    -\frac{k_1}{V} y_{1,1} + k_2( A^t  - y_1 ),\\
\label{app_ma_first_2}
  \partial_t y_{1,1}
  & =
     -6 \frac{k_1}{V} y_1 y_{1,1}  + 4 \frac{k_1}{V}y_1^3 + 2 k_2 (A^t y_1 - y_{1,1}) + \frac{k_1}{V} y_{1,1} + k_2 (A^t-y_1).
\end{align}
Note that these are the resulting second-order MA equations for both the normal \emph{and} the Poisson MA.

\subsection{Applying the conservation laws first}

Alternatively, we can start from the reduced CME with species $B$ and $C$ eliminated, whose stoichiometric matrix and propensity functions are given by
\begin{align}\label{app_stoch_prop_}
  & S
  =
    \begin{pmatrix}
        -1  & 1 
    \end{pmatrix}, \\
  & f_1 (n_1)
   =
    \frac{k_1}{V}n_1^2, \\
  & f_2 (n_1)
  =
    k_2 (A^t-n_1).
\end{align}
Note that due to the term $n_1^2$, the time-evolution equation for the second-order moment $y_{1,1}$ depends on the \emph{diagonal} third-order moment $y_{1,1,1}$ (all moments are diagonal here of course, since we deal with a system with only one variable).
The corresponding equations for the first two moments can be obtained using Eqs.~\eqref{general_eq_first_moment} and \eqref{general_eq_second_moment} and read
\begin{align}
  \partial_t y_1
  & =
    -\frac{k_1}{V} y_{1,1} + k_2( A^t  - y_1 ), \\
\label{app_ma_last_2_nc}
  \partial_t y_{1,1}
  & =
    -2 \frac{k_1}{V} y_{1,1,1} + 2 k_2 (A^t y_1 - y_{1,1}) + \frac{k_1}{V} y_{1,1} + k_2 (A^t-y_1).
\end{align}
For closing these equations to second order, we need to express $y_{1,1,1} $ in terms of $y_1$ and $y_{1,1}$. The corresponding expression is now not the same anymore for the normal and Poisson MAs.  For the normal MA we have $y_{1,1,1} = 3 y_1 y_{1,1} - 2 y_1^3$. Inserting the latter into Eq.~\eqref{app_ma_last_2_nc} one obtains the same result as in Eqs.~\eqref{app_ma_first_1} and \eqref{app_ma_first_2} which we obtained by applying the conservation laws \emph{after} closing the equations. In contrast, if we apply the Poisson MA, which sets $y_{1,1,1} = 3 y_1 y_{1,1} - 2 y_1^3 + y_1$, the resulting equation for $y_{1,1}$ is \emph{not} equal to Eq.~\eqref{app_ma_first_2}. The reason for this is that the Poisson MA does not treat diagonal and non-diagonal moments equivalently. Here, this means that the replacements of $y_{1,1,1}$ and $y_{1,1,2}$ differ from each other if one sets the index $2$ to $1$ in the expression for $y_{1,1,2}$. Since the same is true for the log-normal MA, the latter also gives differing results depending if the equations are closed before or after the conservation laws are applied. Since the normal and CMN-MA do treat diagonal and non-diagonal moments equivalently (so the expressions for $y_{1,1,1}$ and $y_{1,1,2}$ are the same after setting $2$ to $1$), these MAs do not suffer from this flaw.

\end{appendix}
\end{document}